\documentclass[%
superscriptaddress,
 amsmath,amssymb,
 aps,
 prd,
 twocolumn,
floatfix
]{revtex4-2}

\usepackage{graphicx}
\usepackage{amsmath, amssymb}
\usepackage{dcolumn}
\usepackage{bm}
\usepackage{hyperref}
\usepackage{pgfplots}
\usepackage{pgfplotstable}
\usetikzlibrary{arrows,shapes,plotmarks}
\usepackage{xcolor}
\usepackage{wrapfig}
\usepackage{float}
\usepackage{xspace}
\usepackage[caption=false]{subfig}
\usepackage{multirow}

\pgfplotsset{compat=1.3}

\newcommand\aap{\ref@jnl{A\&A}}
\newcommand\aapr{\ref@jnl{A\&A~Rev.}}
\newcommand\apjs{\ref@jnl{ApJS}}
\newcommand\araa{\ref@jnl{ARA\&A}}
\newcommand\mnras{\ref@jnl{MNRAS}}
\newcommand\na{\ref@jnl{New A}}
\newcommand\nar{\ref@jnl{New A Rev.}}

\newcommand{\BSSN}{\textsc{BSSN}}

\newcommand{\dendrogr}{\textsc{Dendro-GR}\xspace}
\newcommand{\dendro}{\textsc{Dendro}\xspace}

\newcommand{\lazev}{\textsc{LazEv}\xspace}

\newcommand{\dgn}{\textit{octant local}}
\newcommand{\cgn}{\textit{octant shared}}

\newcommand{\unzip}{\textit{unzip}}
\newcommand{\zipped}{\textit{zipped}}
\newcommand{\unzipped}{\textit{unzipped}}

\newcommand{\orcidicon}[1]{\href{https://orcid.org/#1}{\includegraphics[height=\fontcharht\font`\B]{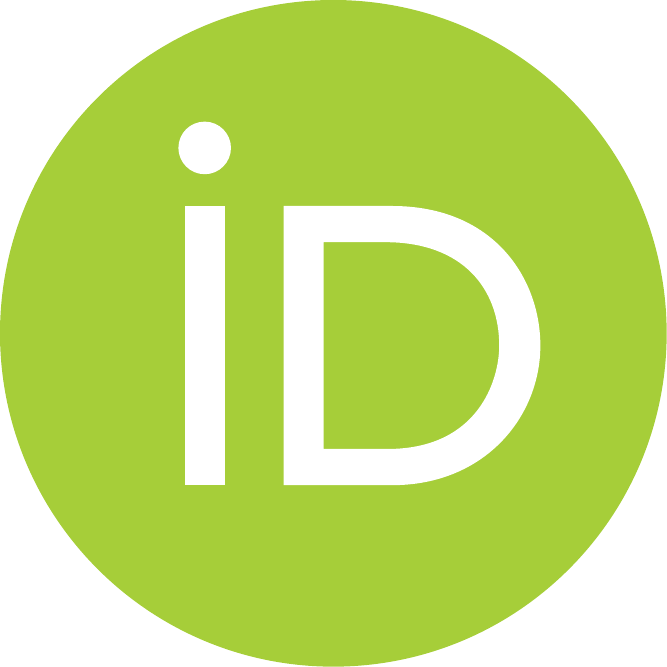}}}

\begin{document}

\title{Massively parallel simulations of binary black holes with Dendro-GR}

\author{Milinda Fernando \orcidicon{0000-0002-5502-9367}}
 \email{milinda@oden.utexas.edu}
 \affiliation{%
 	Oden Institute for Computational Engineering and Sciences, University of Texas at Austin, TX 78712, USA
 }%

\author{David Neilsen \orcidicon{0000-0002-6142-5542}}
 \email{david.neilsen@byu.edu}
\affiliation{
Department of Physics \& Astronomy, Brigham Young University, Provo, UT 84602, USA
}%

\author{Yosef Zlochower \orcidicon{0000-0002-7541-6612}}
 \email{yrzsma@rit.edu}
\affiliation{%
Center for Computational Relativity and Gravitation, and School of Mathematical Sciences, Rochester Institute of Technology,\\
85 Lomb Memorial Drive, Rochester, New York 14623, USA
}%

\author{Eric W.~Hirschmann}
 \email{ehirsch@byu.edu}
\affiliation{
Department of Physics \& Astronomy, Brigham Young University, Provo, UT 84602, USA
}%

\author{Hari Sundar}%
 \email{hari@cs.utah.edu}
\affiliation{%
School of Computing, University of Utah, Salt Lake City, UT 84112, USA
}

\date{\today}

\begin{abstract}
We present results from the new \dendrogr code.  These include simulations 
of binary black hole mergers for mass ratios up to $q=16$.  \dendrogr uses
Wavelet Adaptive Multi-Resolution (WAMR) to generate an unstructured grid 
adapted to the spacetime geometry together with an octree based data structure.
We demonstrate good scaling, improved convergence properties and efficient 
use of computational resources.  We validate the    
code with comparisons to \lazev.  
\end{abstract}


\maketitle


\section{\label{sec:intro}Introduction}
The gravitational wave detectors LIGO/Virgo have made a number of epochal 
discoveries~\cite{theligoscientificcollaboration2021gwtc3,
theligoscientificcollaboration2021population}.  These have given us a dramatically broader conception and 
understanding of the high-energy universe and some of its compact object 
constituents~\cite{Abbott:2016blz,LIGOScientific:2017vwq,PhysRevX.9.011001}
As these detectors continually improve~\cite{PhysRevD.91.062005,aplusdesign,ligovoyager} and are added to by 
new detectors, such as KAGRA~\cite{galaxies10030063}, we can confidently expect an ongoing parade of
additional discoveries.   

The detection and analysis of gravitational waves uses a library of modeled
waveforms for comparison with the detector output signal. Numerical relativity
waveforms are computed using the full nonlinear, Einstein equations, and these
waveforms span the evolution of the binary system from inspiral, through
merger, and finally to ringdown. These waveforms may be used directly in the
analysis of gravitational waves~\cite{Lange:2017wki, LIGOScientific:2016kms}, 
or to inform and validate faster, approximate
methods for generating waveforms, such as semi-analytical and phenomenological 
methods (see, e.g.,~\cite{Pan:2009wj,Pan:2011gk, Taracchini:2012ig, Pan:2013rra, Cotesta:2018fcv, Husa:2015iqa, London:2017bcn, Garcia-Quiros:2020qpx}).  
Numerical relativity
can also probe certain astrophysical scenarios that are difficult to model with
approximate methods.  Examples of such scenarios include non-vacuum 
spacetimes, such as systems with neutron stars, accretion disks, 
and/or magnetic fields.  Even some vacuum binary
black hole systems can be difficult to model with approximate techniques,
such as binaries with large eccentricity, high spins, or large mass
ratios.  
We use the mass ratio $q=m_1/m_2$, 
where $m_1$ is the mass of the primary with $m_1\ge m_2$.

While there are advantages to using numerical
relativity waveforms directly in gravitational wave analysis, there are
significant challenges in calculating waveforms of sufficient quality.
The waveforms must be sufficiently long, have errors bounded within
known tolerances, and they must span a large region of the binary 
parameter space.  The development of newer, more sensitive gravitational wave 
detectors significantly complicates the challenge.  For example,
recent work on requirements for third generation 
detectors~\cite{Punturo:2010zz,Reitze:2019iox,PhysRevD.91.082001}
and LISA~\cite{PhysRevD.76.104018} estimate that errors in numerical relativity waveforms 
need to be reduced by an order of magnitude~\cite{Purrer:2019jcp}.
Another study found that numerical resolutions of
BBH spacetimes will need to be increased by almost a factor of ten in 
some cases~\cite{Ferguson:2020xnm}. 
Reducing the error in numerical waveforms to the level 
required by 3G detectors will require new algorithms and methods in 
numerical relativity.

The challenge of producing waveforms for future gravitational
wave detectors will require highly scalable numerical relativity codes
that are able to efficiently run on exascale supercomputers.
\dendrogr\ is a new code for relativistic astrophysics that is designed to
meet some of the next-generation challenges in numerical relativity.
\dendrogr\ scales well on massively parallel supercomputers, and it uses fast,
responsive adaptive multi-resolution based on wavelets (WAMR). Importantly,
\dendrogr\ easily accommodates many well-tested numerical methods that have been
developed in the relativity community, such as the evolution of Einstein
equations in the \BSSN\ formalism and high-resolution shock-capturing methods
for relativistic fluid dynamics. 

Several projects are currently being developed in the community that use
modern adaptive-mesh infrastructures and sophisticated numerical algorithms to
meet this computational challenge.
Among these are  {\sc GR-Athena++}~\cite{Daszuta:2021ecf}, which
uses the highly-efficient octree AMR infrastructure of {\sc Athena++}
for full numerical relativity simulations coupled to GRMHD, {\sc
GR-Chombo}~\cite{Andrade:2021rbd}, a fully modern AMR numerical
relativity code allowing for complex grid configurations, and {\sc
CarpetX}, which is a new AMR driver for the Einstein
Toolkit~\cite{EinsteinToolkit:2021_11, Loffler:2011ay} that is built
on the {\sc AMReX} toolkit~\cite{amrex:2022}.
Pseudospectral and discontinuous Galerkin methods promise some advantages
for massively parallel computing.  
{\sc Spectre}~\cite{Kidder:2016hev} uses
discontinuous Galerkin methods and a task-based parallelization
scheme. Nmesh~\cite{tichynmesh2020,tichynmesh2022} and bamps~\cite{PhysRevD.93.063006} are other codes 
using DG methods.
{\sc Simflowny}~\cite{Palenzuela_2021} has
a domain specific language and a web-based development environment and 
graphical user interface.  Simflowny
can generate code for multiple platforms, such as SAMRAI~\cite{SAMRAI}.
\dendrogr\ uses an efficient octree structure to store the grid
elements similar in spirit to that used in {\sc GR-Athena++}. While
\dendrogr's wavelet decomposition with an unstructured grid is
similar in spirit to {\sc Spectre}.

This paper presents results from some of the
first binary black hole mergers performed with \dendrogr. We study
gravitational waves from binary black hole systems with 
mass ratios up to $q=16$.  We compare results with 
the well-known \lazev~\cite{Zlochower:2005bj, Campanelli:2005dd} code in some cases, and find that the solutions
match in the convergence limit.  We also present performance 
data for \dendrogr.

\section{\label{sec:meth}Methods}

\dendrogr\ has been built with the intention of tackling relativistic 
astrophysics problems involving merging compact objects.  Its development 
uses and accommodates a number of standard techniques within numerical 
relativity as well as including some new approaches; all with an eye to 
improving the efficiency, scalability 
and time to solution for still challenging problems such 
as large mass ratio binary black holes.  Among the conventional and 
well-tested numerical methods used in \dendrogr\, we solve the Einstein 
equations using the BSSN formulation together with typical coordinate 
conditions, 
initial data, and finite differencing algorithms.  Newer approaches used 
within \dendrogr\ include some of the following and are discussed at greater 
length subsequently in this section.  The code uses a dynamic grid which 
is constructed via an expansion of the grid functions in an interpolating 
wavelet basis.  In this basis, terms in the wavelet expansion can be mapped to 
individual grid points.  The resulting unstructured grid is naturally 
represented computationally as an octree.  On integrating the equations of 
motion in time, each node of this octree is separately \textit{unzipped} 
(decompressed) into a local point representation on a uniform Cartesian 
grid.  The integrated functions are then \textit{zipped} (compressed) back 
to a sparse representation by thresholding the coefficients of the wavelet 
expansion.  This sparse representation is compact and computationally 
efficient as it conserves computer memory and reduces parallel 
communication.   This section describes some of these key components 
of \dendrogr\ in more detail.
We begin with a brief description of our formalism for solving the 
Einstein equations and setting initial data.  We then describe the 
generation of the grid using WAMR and the process for integrating the equations.

\subsection{Formalism}
\label{sec:nm}

There is an extensive literature on solving the \BSSN\ equations in
general relativity, including monographs such 
as~\cite{Alcubierre:1138167,Baumgarte:2010ndz,
Shibata:2015:NR:2904075,Rezzolla_book}.
This section briefly outlines our particular choices for
solving the \BSSN\ equations.
We write the \BSSN\ equations in terms of the conformal
factor~\cite{Campanelli:2005dd} 
\begin{equation}
\chi^{-1} = \det(\gamma_{ij}).
\end{equation}
For gauge conditions, we use the
``$1+\log$'' slicing condition and the $\Gamma$-driver shift as used
in~\cite{Neilsen:2014hha} 
\begin{eqnarray}
\partial_t\alpha & = & \beta^i \partial_i \alpha - 2\alpha K \\ 
\partial_t\beta^i & = & \beta^j \partial_j \beta^i + {3\over4} B^i \\ 
\partial_t B^i & = & \beta^j \partial_j B^i + \partial_t {\tilde\Gamma}^i - \beta^j \partial_j {\tilde \Gamma}^i - \eta B^i .
\end{eqnarray} 
Spatial derivatives are calculated using centered 
finite difference operators that are ${\rm O}(h^6)$ in the grid spacing, $h$. 

The semi-discrete Einstein equations are integrated in time using explicit 
fourth-order Runge-Kutta and a CFL of 0.25. Kreiss-Oliger dissipation
is added to the equations using a fifth-order operator
\begin{align}
\Delta^6_x u^n_m &= \left(-u^n_{m+3} + 6 u^n_{m+2} - 15 u^n_{m+1}+ 20u^n_m\right.\nonumber \\
& \quad \left. -15 u^n_{m+1} + 6 u^n_{m-2} - u^n_{m-3}\right)/\left({64 \triangle x}\right),  
\label{eq:kodiss}
\end{align}
with a tunable amplitude parameter $\sigma$, $0 \le \sigma < 1$, 
which allows one to adjust the amount of dissipation~\cite{Alcubierre:1138167}.  
As discussed below
in Section~\ref{sec:KOdisstest}, we found best results with $\sigma = 0.4$.
We make the common choice to enforce certain algebraic constraints and 
derivative definitions as described, for example, in~\cite{Bruegmann:2006at}.
Outgoing radiative boundary conditions are applied to the dynamical variables. 

We extract gravitational waves from our simulations at five radii 
between $50\: M \le r \le 100\: M$ using
the Penrose scalar, $\psi_4$~\cite{Alcubierre:1138167,Bishop_2016_gw}.
Their decomposition with respect to spin weighted spherical
harmonics (SWSH) is performed using 
Lebedev quadrature~\cite{lebedev1977spherical}. 
To evaluate $\psi_4$ at each of the quadrature points on each 2-sphere, we 
perform an efficient search operation on the underlying grid, and SWSH projection coefficients 
are computed with a parallel
reduction operation~\cite{fernando2019massively}.

\subsection{Initial Data}
\label{sec:id}

Initial data for both \dendrogr\ and \lazev\ are set using the \textsc{TwoPunctures} 
code~\cite{Ansorg:2004ds}
from the {\sc EinsteinToolkit}~\cite{Loffler:2011ay, EinsteinToolkit:2021_11}. 
For the initial values of shift, both codes set
$\beta^i(t=0) =0$.  Initial values of the lapse in \dendrogr\ 
use the ad-hoc function
$\alpha(t=0) = \tilde \psi^{-2}$, where
$\tilde \psi = 1 + m_{p1}/(2 r_1) + {m}_{p2}/(2 r_2)$,
$r_i$ is the coordinate distance to the $i$th BH, and ${m}_{pi}$ is the bare
mass parameter of the $i$th BH.  
In \lazev\, the initial lapse is 
$\tilde \psi = 1 + 1/(4 r_1) + 1/(4 r_2)$.

\begin{table*}
\caption{The initial configuration parameters for non-spinning
binary black hole systems for increasing mass ratio. The presented
numerical waveforms are based on these initial data parameters. The
parameters for $q\ge 2$ were obtained using the expressions 
in~\cite{Healy:2017zqj}.  The initial data are set using the 
bare mass parameters.
The black holes are placed initially on the $x$-axis at the locations
$x_1$ and $x_2$ as given in the table. The linear momentum of the
the second black hole is given in the last two columns, $\mathbf{p}_2 = (p_x,p_y)$,
and $\mathbf{p}_1 = - \mathbf{p}_2$.  \label{tab:tpid}}
\begin{ruledtabular}
\begin{tabular}{cccccccccc}
Mass ratio & \multicolumn{2}{c}{Puncture Parameter} & \multicolumn{2}{c}{ADM Mass} & Total& \multicolumn{2}{c}{x-position} & \multicolumn{2}{c}{Momentum BH2}\\
$q=m_1/m_2$ & \multicolumn{1}{c}{$m_{p2}$} & \multicolumn{1}{c}{$m_{p1}$} & \multicolumn{1}{c}{$m_2$} & \multicolumn{1}{c}{$m_1$} &
ADM Mass & \multicolumn{1}{c}{$x_2$} & \multicolumn{1}{c}{$x_1$} & \multicolumn{1}{c}{$p_x$} & \multicolumn{1}{c}{$p_y$} \\ 
\hline
1 & 4.8240E-1   & 4.8240E-1 & 5.0010E-1 & 5.0010E-1 & 9.8844E-1 & 4 & -4 & 0 &  0.1140 \\
2 & 0.31715   & 0.65150 & 6.6667E-1 & -3.3333E-1 & 9.8931E-1    & 5.3238 & -2.6762 & -1.7777E-03 & 1.0049E-01 \\
4 & 1.8805E-01 & 7.8937E-01 & 2.0000E-1 & 8.0000E-1 & 9.9237E-1 & 6.3873 & -1.6127 & -1.0647E-03 & 7.2660E-02 \\
8 & 1.0362E-01 & 8.8245E-01 & 1.1111E-1 & 8.8889E-1 & 9.9534E-1 & 7.1006 & -8.9938E-1 & -4.9937E-04 & 4.5037E-02 \\
16 & 5.4585E-02 & 9.3761E-01 & 5.8824E-2 & 9.4118E-1 & 9.9740E-1 & 7.5226 & -4.7741E-1 & -1.9532E-04 & 2.5319E-02 
\end{tabular}
\end{ruledtabular}
\end{table*}

In this paper, we evolve non-spinning black hole binaries with mass ratios 
$q=1, 2, 4, 8$, and $16$.
We place the black holes initially on the
$x$-axis, with the binary's center of mass at the origin.  
The smaller black hole with mass $m_2$ is
placed on the positive $x$-axis, and the initial coordinate separation is 
fixed to $x_2 - x_1 = 8\: M$.
Initial data parameters for the $q=1$ binary are ad-hoc quasi-circular 
parameters, chosen to match previous work~\cite{Neilsen:2010ax}. 
Parameters for all other cases were found using the low eccentricity 
post-Newtonian expressions reported in~\cite{Healy:2017zqj}. 
To simplify comparisons with \lazev, we set the \textsc{TwoPunctures} code 
to use the bare puncture masses and other parameters shown in 
Table~\ref{tab:tpid} directly.  
Finally, we ran the $q=1, 2$, and $4$ cases with both \dendrogr\ and \lazev, while the 
higher mass ratio simulations were only run with \dendrogr.

\subsection{Symbolic code generation}
\label{subsec:sympygr}

The evaluation of the \BSSN\ equations at a given grid point is 
computationally expensive and can be challenging due to the large number of 
terms associated with the equations. Manually writing code to evaluate 
these equations can be prone to error, difficult to debug, and 
challenging to perform architecture
specific optimizations. To address some of these issues, we have developed a 
SymPy-based code generation framework for 
\dendrogr.  This tool has some of the same capabilities as NrPy+~\cite{nrpy,PhysRevD.97.064036}, but is 
more limited in scope.
Using our symbolic framework, we compute the directed 
acyclic graph representing the underlying computations for the 
\BSSN\ equations. We perform optimizations to reduce the overall number of 
operations as well as architecture specific optimizations that improve our 
code's performance portability. The current implementation of the symbolic 
framework supports CPUs and GPUs~\cite{fernando2019scalable,fernando2019massively}.  

\subsection{Grid generation with WAMR}
\label{subsec:comp_grid}

The computational complexity of the 
Einstein equations, together with the 
requirement of high accuracy across multiple spatial and temporal 
scales, motivates the use of grid adaptivity.  
\dendrogr\ uses a wavelet-based approach which results in a representation 
of the underlying field 
variables on a sparse, adaptive mesh.  We describe briefly here the fundamental 
aspects of this sparse representation.  More complete details can be found 
in~\cite{debuhr2015relativistic,fernando2017machine,fernando2019massively,fernando2019scalable}. 
While we use the coefficients in a wavelet expansion to generate the 
computational grid, we store the grid functions only in the point 
representation. Thus, the wavelet coefficients are not used to integrate
the equations of motion.

\begin{figure}[tb]
    \centering
        \resizebox{1.0\columnwidth}{!}{
        \begin{tikzpicture}[yscale=-1.2]
            \begin{scope}
                    
                \node[text width=1cm] at (-1.25,0) {{\large (a)} \;\;\quad};
                
                \foreach \y in {0,1,2}
                \node[text width=1cm] at (-0.4,\y) {$V_\y$};
                \foreach \x in {0,4,8,12,16}
                \fill (\x,0) circle (0.1cm);
                
                \foreach \x in {0,2,4,6,8,10,12,14,16}
                \fill (\x,1) circle (0.1cm);		
                
                \foreach \x in {0,1,2,3,4,5,6,7,8,9,10,11,12,13,14,15,16}
                \fill (\x,2) circle (0.1cm);
            \end{scope}
            \label{fig:grid_construction:a}
        \end{tikzpicture}}
        \vspace{2pt}
        
        \resizebox{1.0\columnwidth}{!}{            
        \begin{tikzpicture}[yscale=-1.2] 
            \begin{scope}[yshift=4cm]
                
                \node[text width=1cm] at (-1.25,0) {{\large (b)} \;\;\quad};
                
                \node[text width=1cm] at (-0.4,0) {\small $V_0$};
                \foreach \y in {1,2}
                \node[text width=1cm] at (-0.4,\y) {\small $W_\y$};
                \foreach \x in {0,4,8,12,16}
                \fill (\x,0) circle (0.1cm);
                
                \foreach \x in {2,6,10,14}
                \fill[red] (\x,1) circle (0.1cm);		
                
                \foreach \x in {1,3,5,7,9,11,13,15}
                \fill[blue] (\x,2) circle (0.1cm);
            \end{scope}    
            \label{fig:grid_construction:b}
        \end{tikzpicture}}
        \vspace{2pt}
        
        \resizebox{1.0\columnwidth}{!}{
        \begin{tikzpicture}[yscale=-1.2]
            \begin{scope}[yshift=8cm]
                
                \node[text width=1cm] at (-1.25,0) {{\large (c)} \;\;\quad};
                
                \node[text width=1cm] at (-0.4,0) {\small $V_0$};
                \foreach \y in {1,2}
                \node[text width=1cm] at (-0.4,\y) {\small $W_\y$};
                \foreach \x in {0,4,8,12,16}
                \fill (\x,0) circle (0.1cm);
              
                \foreach \x in {2,6,10,14}
                \fill[red] (\x,1) circle (0.1cm);		
              
                \draw[red] (2,1) circle (0.2cm);
                \draw[red] (6,1) circle (0.2cm);
                \draw[red] (14,1) circle (0.2cm);
              
                \foreach \x in {1,3,5,7,9,11,13,15}
                \fill[blue] (\x,2) circle (0.1cm);
              
                \draw[blue] (1,2) circle (0.2cm);
                \draw[blue] (3,2) circle (0.2cm);
                \draw[blue] (5,2) circle (0.2cm);
                \draw[blue] (13,2) circle (0.2cm);
                \draw[blue] (15,2) circle (0.2cm);
            \end{scope}    
            \label{fig:grid_construction:c}
        \end{tikzpicture}}
        \vspace{2pt}
        
        \resizebox{1.0\columnwidth}{!}{
        \begin{tikzpicture}[yscale=-1.2]
            \begin{scope}[yshift=12cm]
		
                \node[text width=1cm] at (-1.25,0) {{\large (d)} \;\;\quad};
		
                \node[text width=1cm] at (-0.4,0) {\small $V_0$};
                \foreach \y in {1,2}
                \node[text width=1cm] at (-0.4,\y) {\small $W_\y$};
                \foreach \x in {0,4,8,12,16}
                \fill (\x,0) circle (0.1cm);
                
                \foreach \x in {2,6,14}
                \fill[red] (\x,1) circle (0.1cm);		
                
                \draw[red] (2,1) circle (0.2cm);
                \draw[red] (6,1) circle (0.2cm);
                \draw[red] (14,1) circle (0.2cm);
                
                \foreach \x in {1,3,5,13,15}
                \fill[blue] (\x,2) circle (0.1cm);
              
                \draw[blue] (1,2) circle (0.2cm);
                \draw[blue] (3,2) circle (0.2cm);
                \draw[blue] (5,2) circle (0.2cm);
                \draw[blue] (13,2) circle (0.2cm);
                \draw[blue] (15,2) circle (0.2cm);
              
            \end{scope}
            \label{fig:grid_construction:d}
        \end{tikzpicture}}
\caption{\small This illustrates a simplified conception of how the grid is 
constructed.  In {\bf (a)}, one dimensional nested grids, $V_j$, (for 
$j=0,1,2$) are shown.  Note that every fine grid contains all the gridpoints
on every coarser grid.  This redundancy is removed in {\bf (b)} on defining 
the complementary spaces, $W_j$ (colored red and blue).  We compute wavelet 
coefficients ($d^{j,k}$) as the difference between $u^{j,k}$ and the 
field as interpolated from level $V_{j-1}$.  In {\bf (c)}, those gridpoints 
with wavelet coefficients larger than a predetermined threshold, $\epsilon$, 
are tagged (here with circles) as essential to the calculation.  For those 
gridpoints with $|d^{j,k}|<\epsilon$, the corresponding terms in the 
interpolating wavelet expansion are ignored and the gridpoints are discarded 
from the mesh, as illustrated in ${\bf (d)}$.  
\label{fig:grid_construction}}
\end{figure}
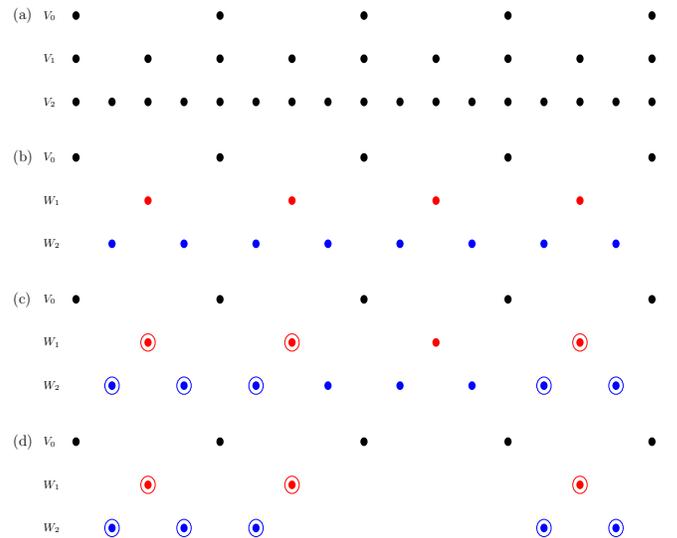

Two essential ingredients in our approach are the notion of iterative 
interpolation~\cite{Deslauriers1989} and the wavelet 
representation itself~\cite{Donoho1992,Holmstrom1999}.  We demonstrate both
of these in one 
dimension. The extension to multiple dimensions is straightforward
and is accomplished 
by simply repeating the procedures we will describe in each additional dimension.  
To fix ideas, we first define a set of nested grids, $V_j$, where 
$$
V_j = \bigl\{ x^{j,k} : x^{j,k} = 2^{-j} \, k \Delta x \bigr\},
$$
where $j$ and $k$ are nonnegative integers and $\Delta x$ is the spacing on 
the base grid (or level) and which is labeled with $j=0$.  This base grid, 
$V_0$, is comprised of $N+1$ gridpoints evenly spaced on a domain of length 
$L=N \Delta x$.  Each finer grid with $j > 0$ contains each point 
in every coarser grid.  Values of a field, $u$, at level $j$ are designated 
$u(x^{j,k}) \equiv u^{j,k}$. If known, these values are copied from coarser 
grids to all required fine grids.  For example to go from $V_j$ to $V_{j+1}$, 
we take $u^{j+1,2k} = u^{j,k}$.  A similar copy happens to all higher level 
fine grids.  Of course, on these finer grids, there will be points newly 
appearing.  The field values on those gridpoints 
new to grid $V_{j+1}$ are interpolated from the known values on the coarser
grid $V_j$.  We generally use Lagrange 
interpolation. In this manner, all fields at any refinement level can be had 
(see Fig.~\ref{fig:grid_construction}). 
This iterated interpolation, continued to arbitrarily large levels 
produces continuous functions with compact support~\cite{Donoho1992}.  

\begin{figure} 
\includegraphics[width=7.cm,angle=0]{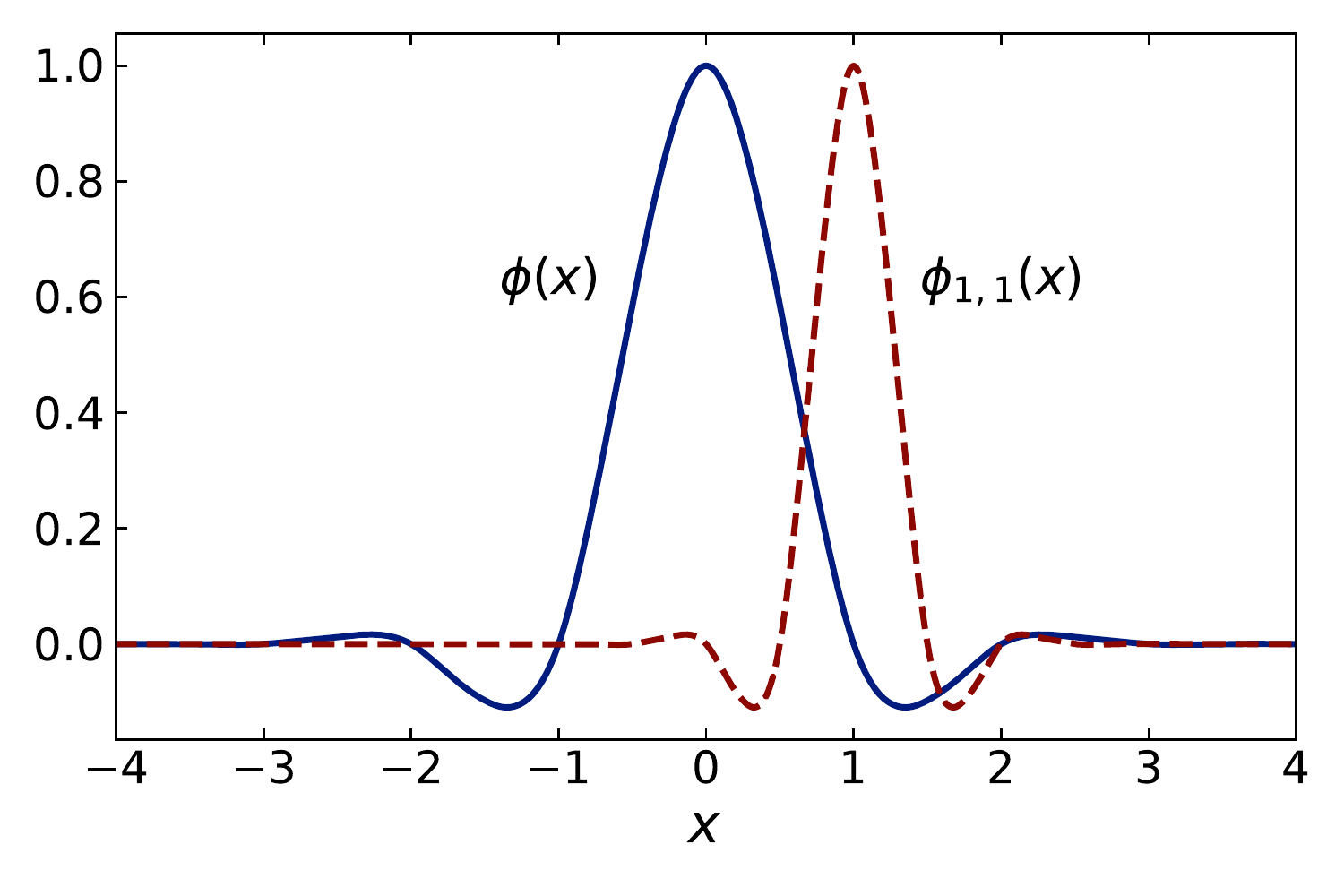}
\caption{This shows the fundamental solution of the iterated interpolation, 
$\phi(x)$ (solid), and a basis element attached to the grid $V_1$ called 
$\phi_{1,1}(x)$ (dashed).  All the basis functions are scaled, translated 
versions of the fundamental solution.} 
\label{fig:basis_fundtion} 
\end{figure} 

The process just described is the start of how our sparse grid will emerge. 
But it also produces a natural 
basis set with which we can represent our fields.  This basis set is 
comprised of interpolating functions created via iteration from a sequence 
of zeros and a single value of one living on $V_0$ (sometimes referred to 
as a Kronecker sequence).  More specifically, define a function $\phi_{0,k}(x)$
which takes values at the points $x^{0,l}$ (imagine on the base grid) of 
$\phi_{0,k}(x^{0,l}) = \delta^l_k$.  Now interpolate as described above to 
find $\phi_{0,k}$ at other gridpoints and iterate.  This can be repeated 
for $\phi_{j,k}$ with $j>0$.  The resulting iterated 
interpolating functions will have a number of properties, including compact
support and a two scale relation given by 
$$
\phi_{j,k}(x) = \sum_l c_{j,k}^l \, \phi_{j+1,l}(x),
$$
where the coefficients 
$c_{j,k}^l$ will depend on the order of the interpolation.
Significantly, each of these iterated, interpolating functions are scaled, 
translated 
versions of a single, fundamental function, $\phi(x)$, 
related to a particular limiting function of the above iterated interpolation.  It 
is related to the Daubechies scaling function and shown in Fig.~\ref{fig:basis_fundtion}.

With these iterated, interpolating functions in hand, we can now return to 
and complete our wavelet representation.  Note that at each point of each 
level, we have an associated scaling function 
$$
\phi_{j,k}(x) = \phi\bigl(2^j x/\Delta x - k\bigr),
$$
which, when taken all together, constructs a basis at each level, $j$.  
However, 
across levels, the set of scaling functions is overdetermined and will not 
form a basis for the entire grid until we deal with the redundancy introduced 
by having common points in $V_j$ and $V_{j+1}$.  To this end, we consider 
the complementary space to $V_j$, which we call $W_j$, such that 
$$
W_j = \bigl\{ x^{j,k} : x^{j,k} = 2^{-j} k \, \Delta x, k \, {\rm odd} \bigr\},
$$
and is that set of points in $V_j$ that are not in $V_{j-1}$ (see Fig.~\ref{fig:grid_construction}b).  With this 
definition, we use the set of grids given by $\{V_0, W_j\}$ and thereby have a 
basis with respect to which we can define our fields $u$:   
$$
u(x) = \sum_{k\in S_0} u^{0,k} \, \phi_{0,k}(x) + \sum_{j=1}^\infty \sum_{k\in S_j} d^{j,k} \, \phi_{j,k}(x) .  
$$
The coefficients $u^{0,k}$ and $d^{j,k}$ are expansion coefficients with 
$S_0 = {0,1, ... , N}$ providing the index set for the base grid, $V_0$, and 
$S_j = {1,3, ... , 2^{j+1} N - 1}$ being the index set for the fine grid given 
by $W_j$.  This last expression is our interpolating wavelet expansion in 
which $u^{0,k}$ are just the values of the field on the base points and the 
coefficients $d^{j,k}$, referred to as wavelet coefficients, are but the 
differences between the field values $u^{j,k}$ and the interpolated values
at $x^{j,k}$ coming from the next lower level, $j-1$.  If we designate 
these interpolated values as ${\tilde u}^{j,k}$ 
the wavelet coefficients are then computed simply as 
$$
d^{j,k} = u^{j,k} - {\tilde u}^{j,k} . 
$$

We can think of that part of the expansion with the scaling 
functions as encoding the smooth part of the field, $u(x)$, while the wavelet 
coefficients provide information about the function on fine scales.  Because
of the highly local nature of the wavelets used, this representation will have
many wavelets in regions exhibiting strong spatial variations while few will 
be necessary in regions where the field is changing slowly.  

With the wavelet representation in hand, compression is now possible.  
More particularly, we can make the representation sparse  
by choosing a threshold value, $\epsilon$, such that if the 
magnitude of the wavelet coefficients, $|d^{j,k}|$, is smaller than $\epsilon$, 
we truncate the expansion and discard the corresponding gridpoints from the grid
itself.  Doing so both reduces the grid size and provides an error bound on 
the representation of the field.  In Fig.~\ref{fig:grid_construction}c and~\ref{fig:grid_construction}d we 
illustrate this approach to constructing the grid.   
As already mentioned, extending to multiple dimensions amounts to taking the basis functions to be
products of the one dimensional basis functions. 

\begin{figure*}[t]
    \begin{center}
    \hbox to \hsize{
    \includegraphics[width=4.45cm]{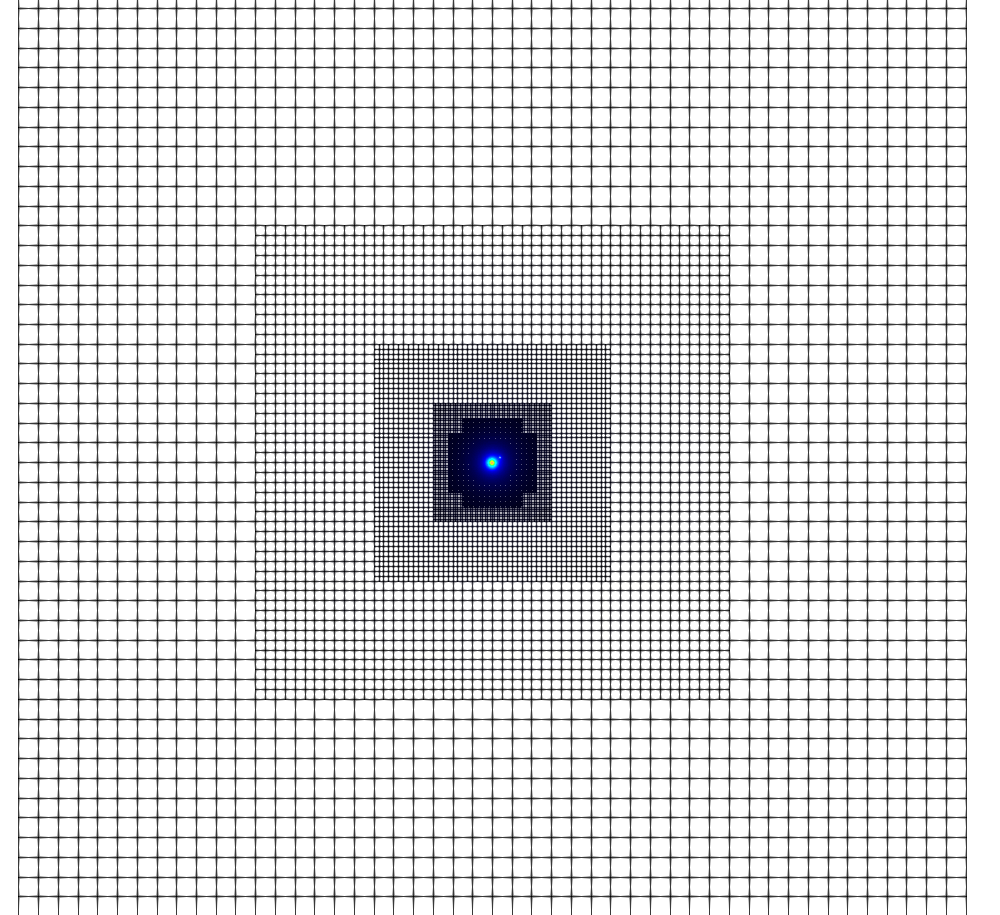}\hfil
    \includegraphics[width=4.45cm]{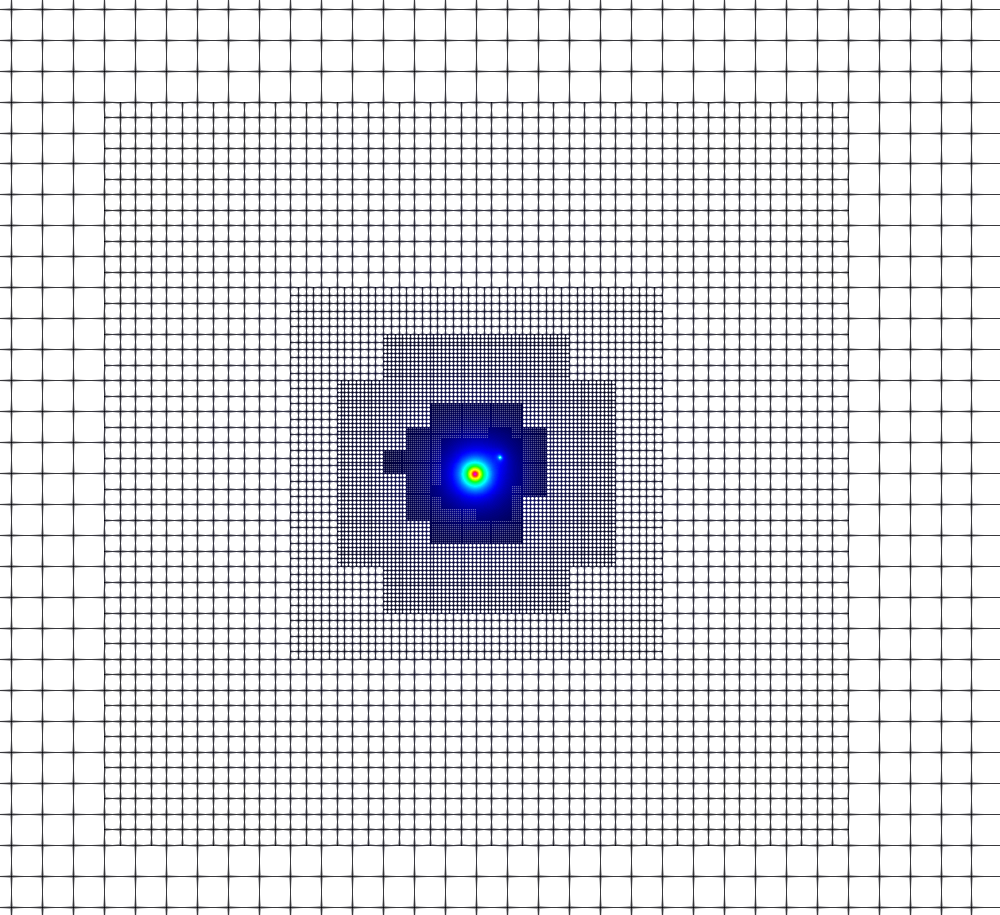}\hfil
    \includegraphics[width=4.45cm]{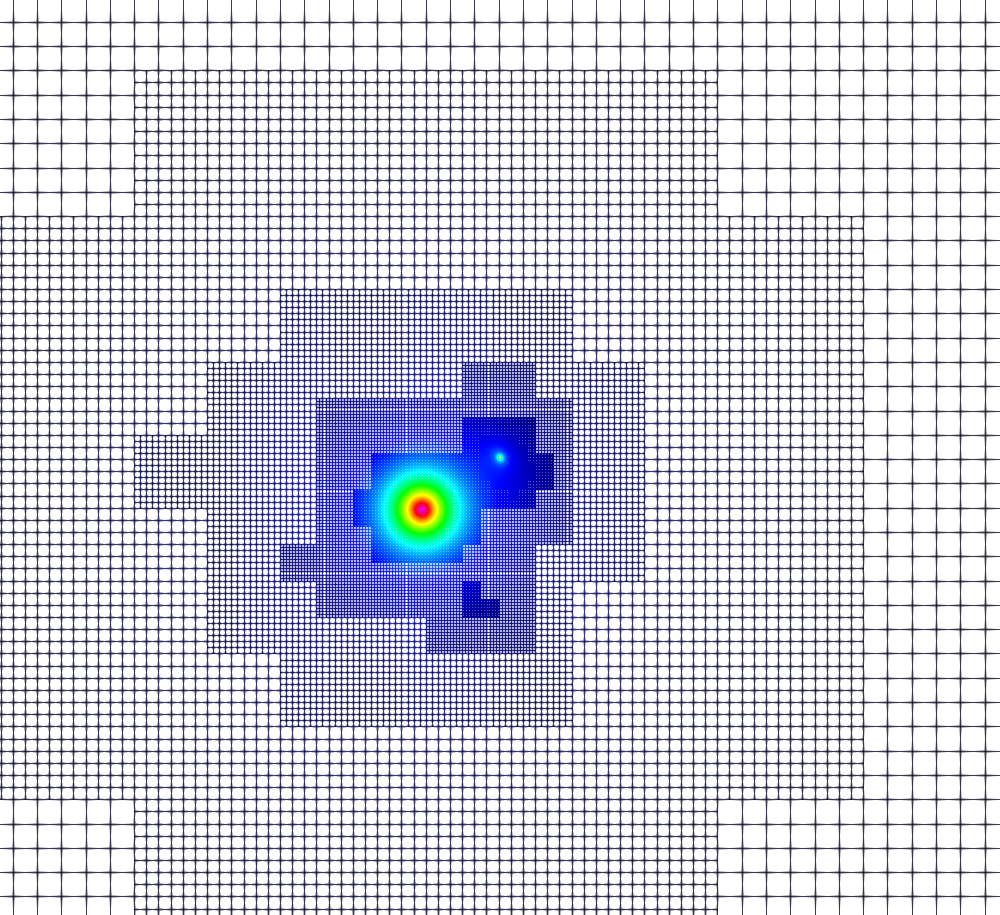}\hfil
    \includegraphics[width=4.45cm]{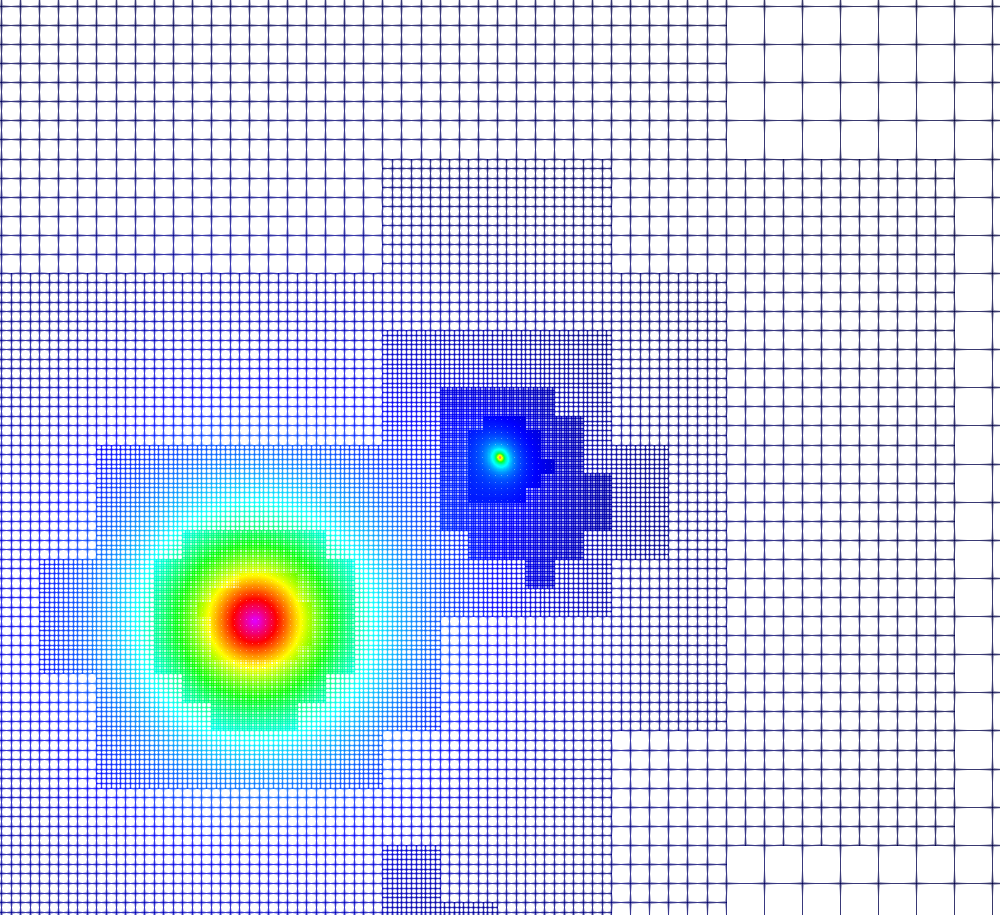}}

    \hbox to \hsize{
    \includegraphics[width=4.45cm]{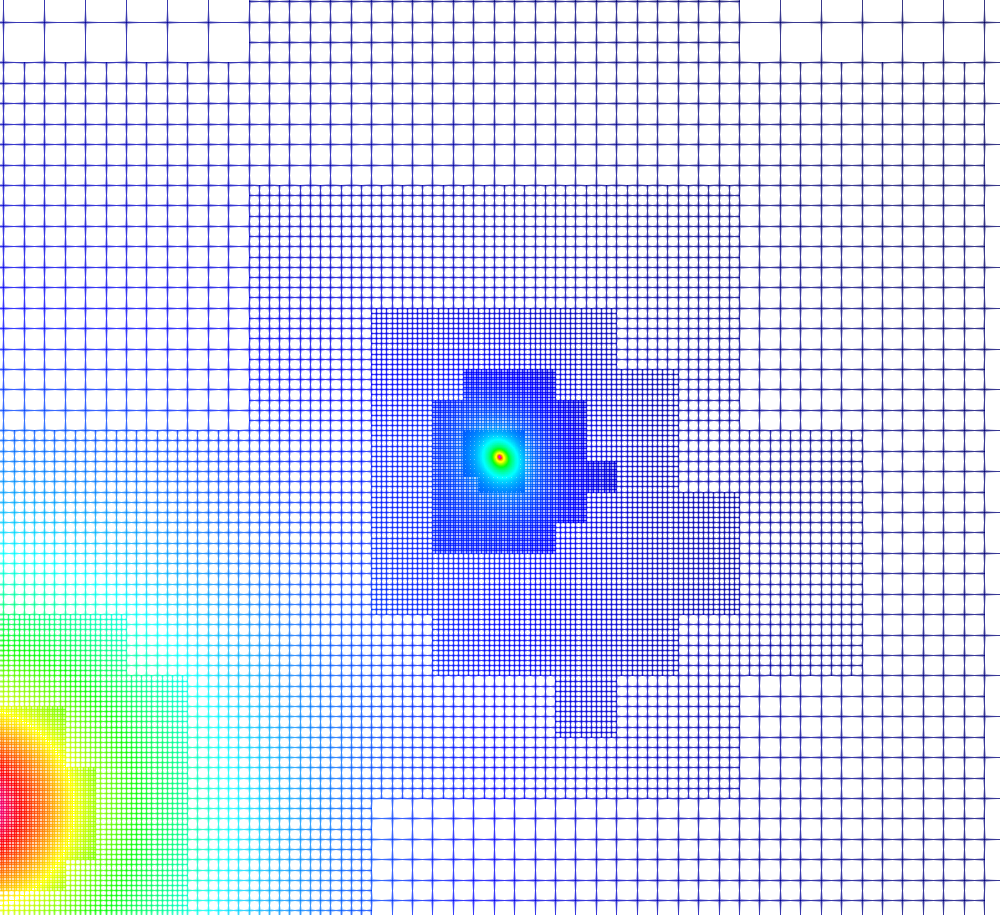}\hfil
    \includegraphics[width=4.45cm]{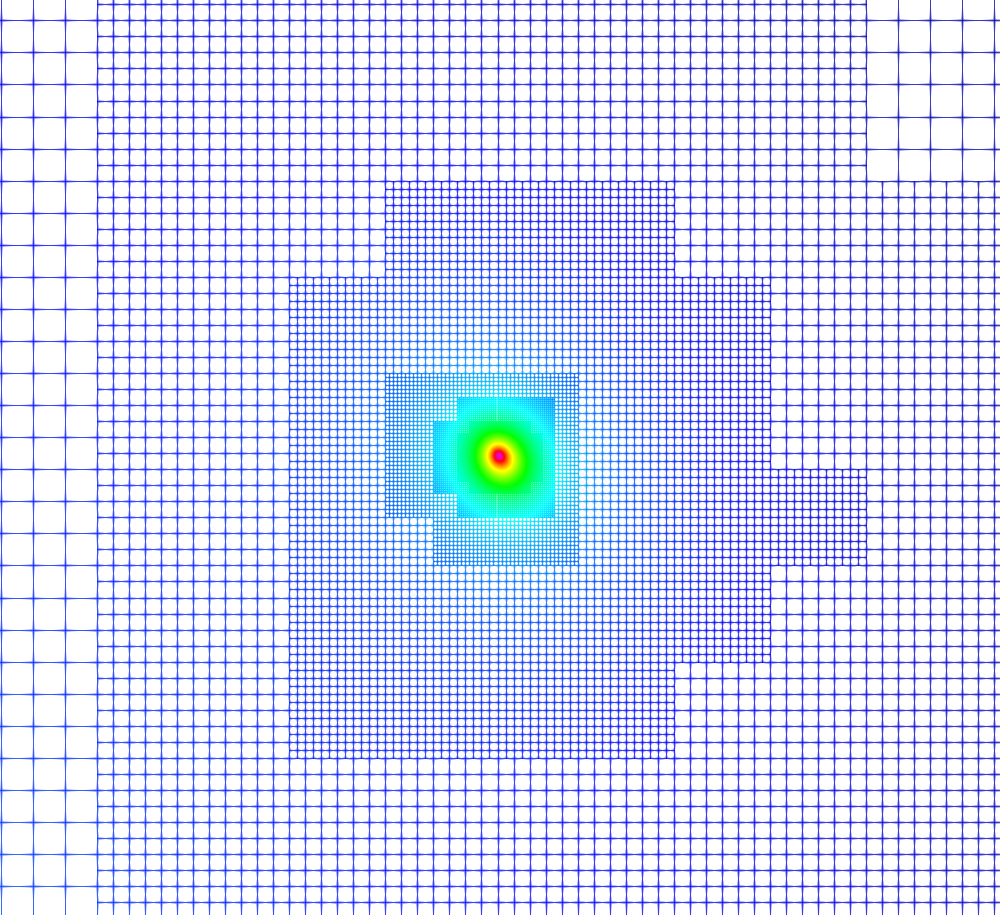}\hfil
    \includegraphics[width=4.45cm]{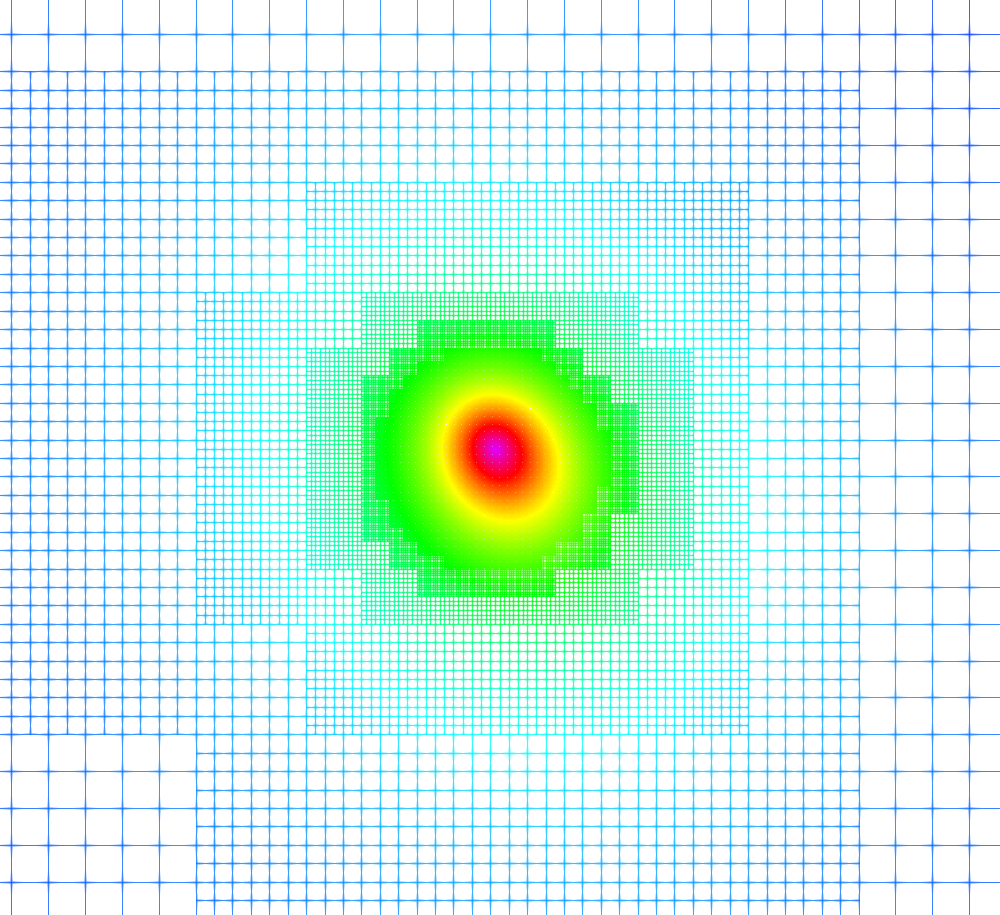}\hfil
    \includegraphics[width=4.45cm]{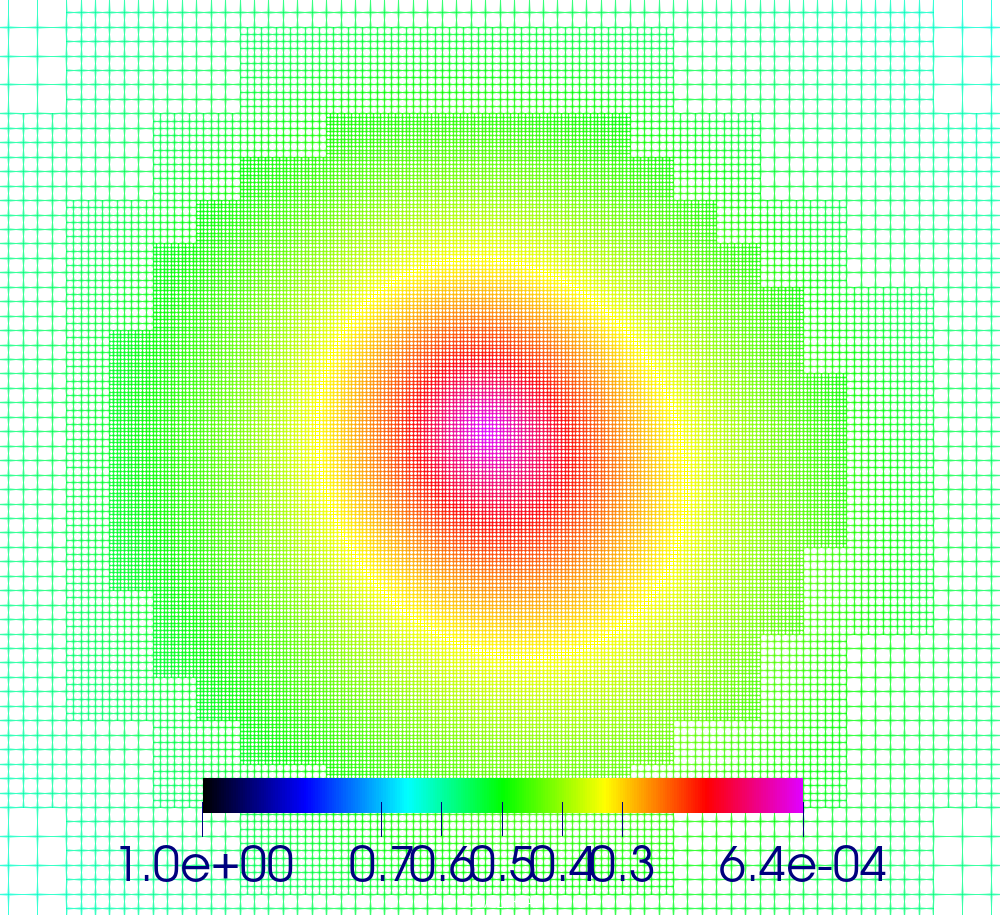}}

    \caption{\small This figure plots the lapse on the computational grid 
generated for a $q=16$ black hole binary, after the system has evolved
for two orbits.  
The top left frame shows the entire 
computational domain, and moving to the right each frame successively 
zooms in towards the smaller of the two black holes.  The computational
grid is sparse, with refinement concentrated about the black holes, making
it very computationally efficient.  The 2:1 refinement constraint for
constructing the grid, discussed in Sec.~\ref{sec:octree_balance},
 is also apparent in the overall grid structure.
\label{fig:wire_frame_grid}
}
\end{center}
\end{figure*}

An example of the WAMR-constructed grid used to evolve binary black holes is
shown in Fig.~\ref{fig:wire_frame_grid}, which shows the grid for a $q=16$ binary 
before merger.  This computational grid is very efficient: the grid is sparse
with refined regions that adapt to the small-scale features of the spacetime.  
The grid does not require refined regions to be rectangular on large scales,
significantly saving on computational and memory costs.  Moreover, large overlapping
regions between refinement levels are not required.

\subsection{Refinement Functions}
\label{sec:rf}

\begin{figure}
\begin{center}
    \includegraphics[width=8.5cm]{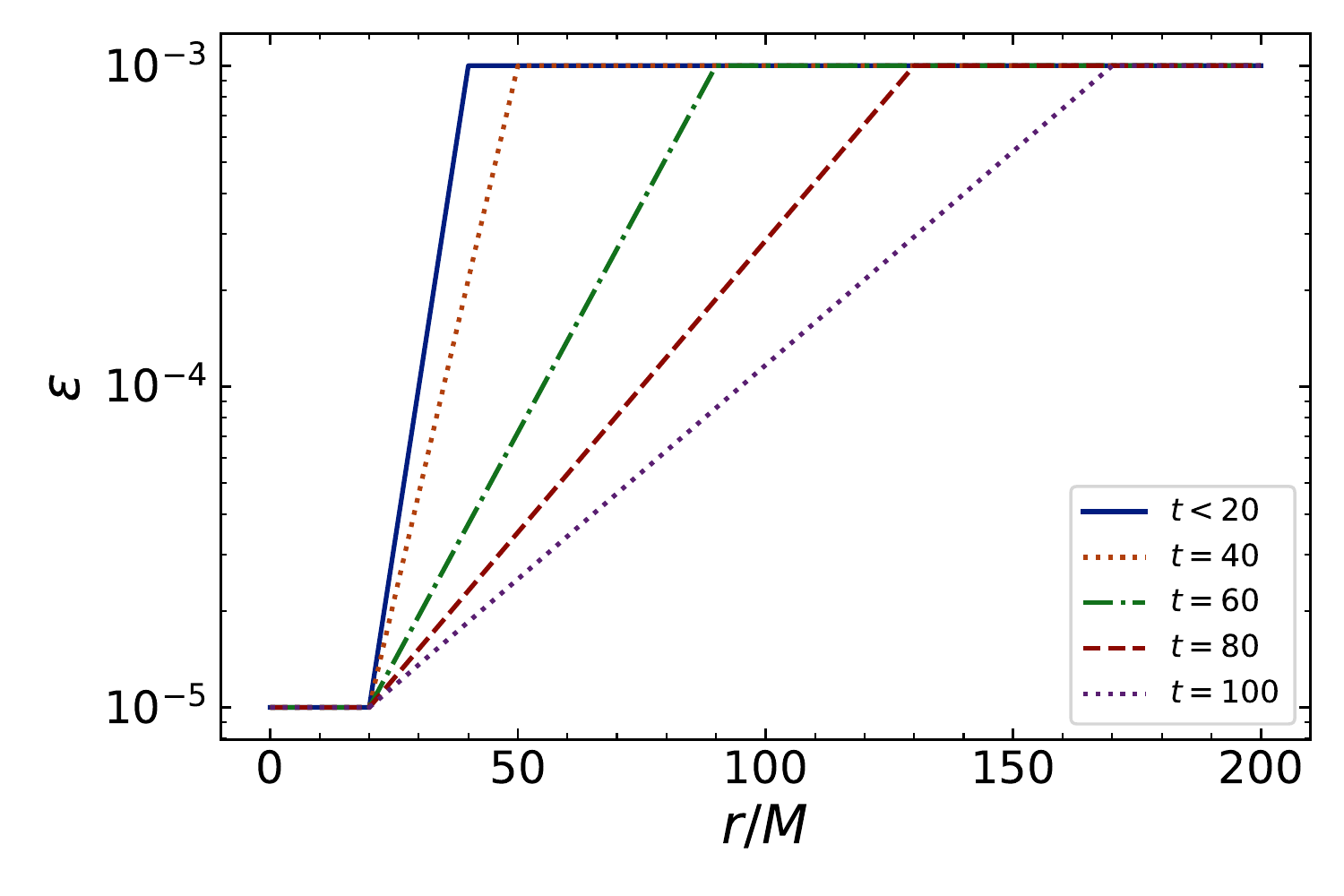}
    \caption{\label{fig:rf3} \small This figure shows the wavelet
tolerance, $\epsilon(r)$, for refinement function RF3 at
a few representative times.  This refinement function is spherically symmetric,
centered on the origin of the grid,
and is independent of the black hole masses.
The minimum wavelet tolerance is used over a relatively large region 
at the center of the grid.  After $t=20M$, the wavelet tolerance decreases 
in the GW extraction zone, $50 < r/M < 100$, 
allowing the initial junk radiation to pass before triggering
refinement in this region.}
\end{center}
\end{figure}

\begin{figure}
\begin{center}
    \hspace*{-0.75cm}\includegraphics[width=5.2cm]{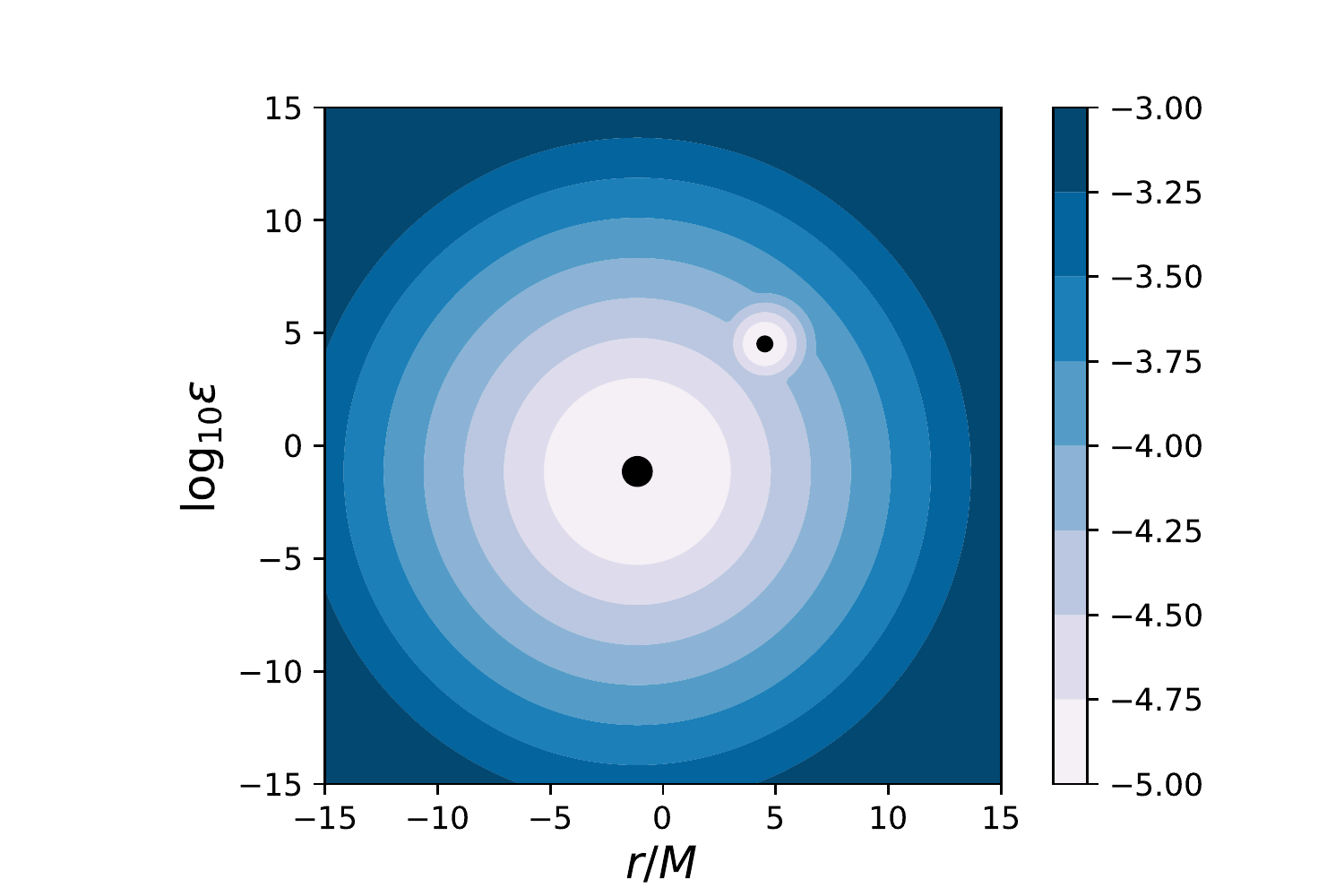}
    \hspace*{-1.10cm}\includegraphics[width=5.2cm]{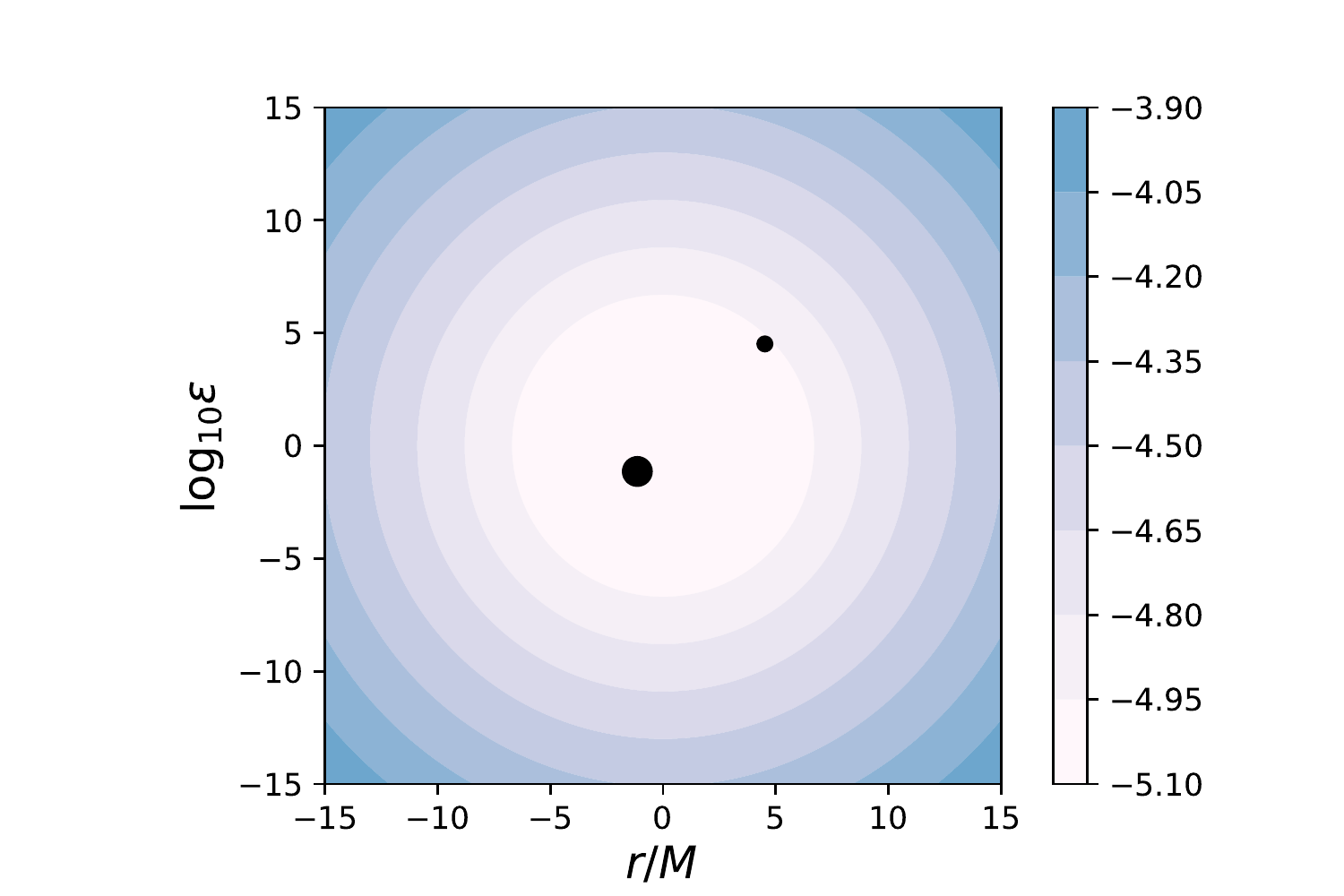}
    \caption{\label{fig:rf4} \small This figure shows the wavelet
tolerance, $\epsilon(r)$, using refinement function RF4 for a 
$q=4$ binary at two times, $t=0$ and $t=40 M$.  
The black dots indicate representative 
positions and relative coordinate sizes for the two black holes, 
though not necessarily the physical horizons.  
This refinement function has the minimum tolerance 
centered about each black hole.  After $t=20M$, the refinement function 
becomes spherically symmetric and centered at the grid origin, with $\epsilon$ 
decreasing in the wave extraction region, similar to RF3.}
\end{center}
\end{figure}

\begin{figure}
\begin{center}
    \hspace*{-0.75cm}\includegraphics[width=5.2cm]{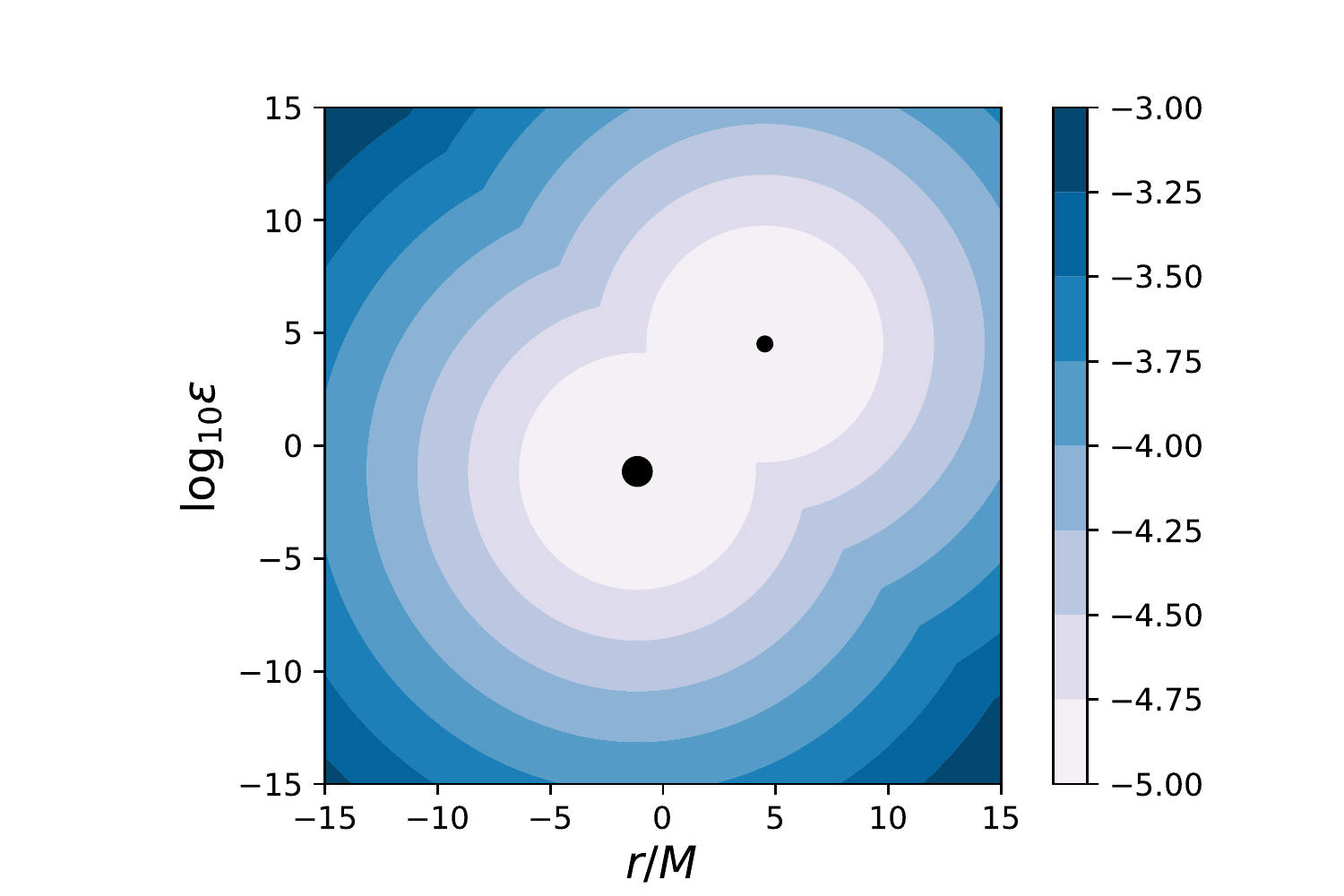}
    \hspace*{-1.10cm}\includegraphics[width=5.2cm]{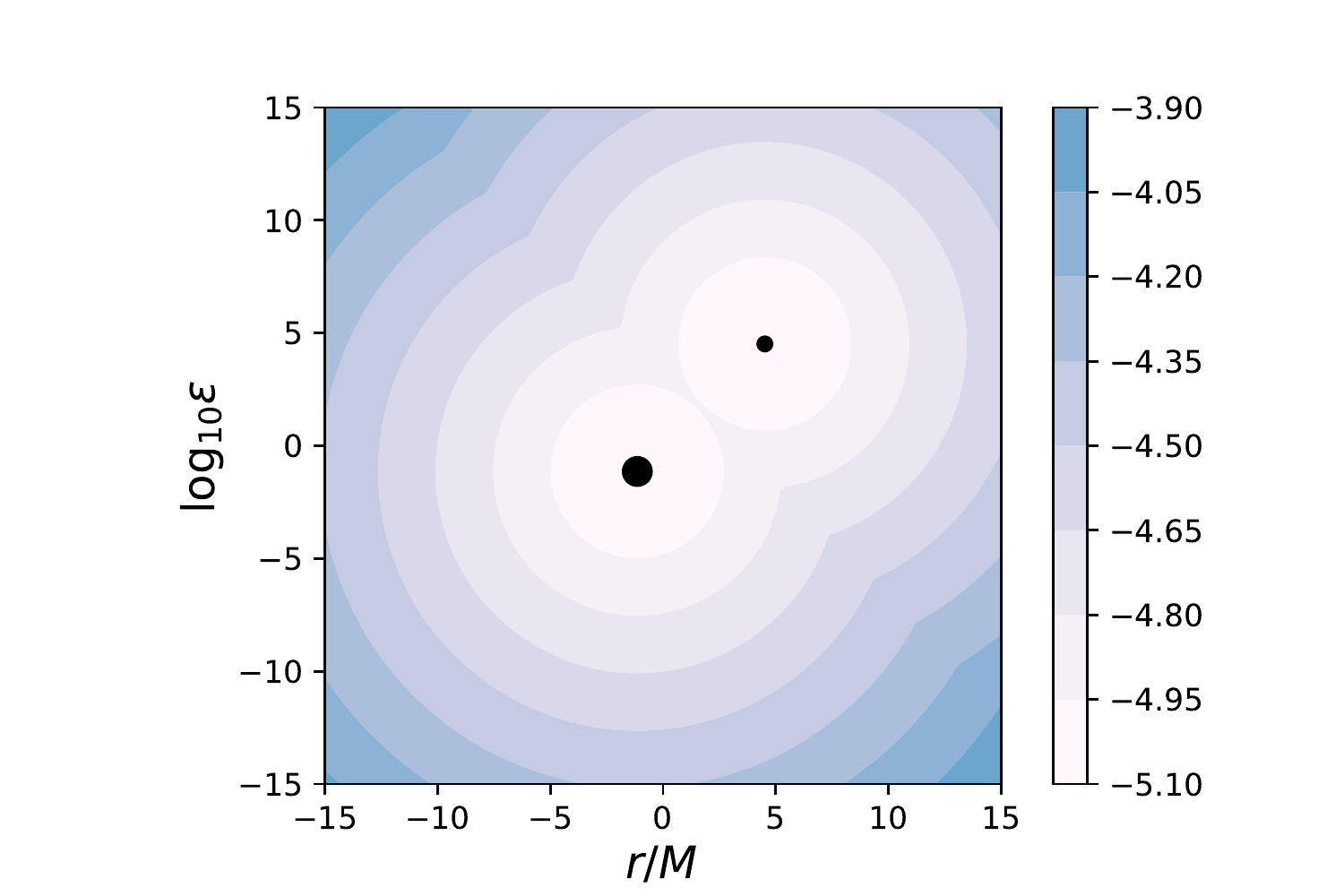}
    \caption{\label{fig:rf5}  \small This figure shows the wavelet
tolerance, $\epsilon(r)$, using refinement function RF5 for a 
$q=4$ binary at two times, $t=0$ and $t=40 M$.
The black dots indicate representative 
positions and relative sizes for the two black holes, however, not the 
physical horizons.  
This refinement function has the minimum tolerance 
centered about each black hole.  After $t=20M$, the refinement function 
reduces to a functional form similar to RF3, but centered about each black 
hole, rather than the grid's origin. }
\end{center}
\end{figure}

Interpolating wavelets are sensitive to any non-differentiable or
non-convergent parts of a solution, triggering immediate refinement.
This is important for resolving small-scale features in solutions.
However, refinement can be triggered by uninteresting or
unphysical features as well.  In binary black hole spacetimes, we
are primarily interested in resolving the binary at the center of
the grid and following radiation out to the extraction region.  To
achieve computational efficiency, therefore, it is important to
control where refinement occurs, focusing on physically interesting
features in the solution.

One way to manage refinement in \dendrogr\ is to set a 
maximum allowed level
of refinement for the entire grid, $J_{\rm max}$.  
This limit is enforced globally
for all times, and is chosen to allow for an expected minimum grid
resolution.  
It is important to note that the spacetime at the black hole puncture is not
smooth, and the WAMR grid will continue refining on this feature until
the maximum level of refinement is reached.  As we evolve binaries
with large mass ratios, we need to prevent over-refinement
of the more massive black hole in the binary.  
We modify the naive use of $J_{\rm max}$ by
tracking the black hole locations and imposing a mass dependent
constraint on the maximum refinement level about each black hole.
We refine a sphere expected to extend
beyond the apparent horizon to the local maximum refinement level.

Recall that refinement in WAMR is controlled by the wavelet 
tolerance $\epsilon$.  Usually, $\epsilon$ is taken to be a constant.
However, we have found that using
a spatially-dependent wavelet tolerance, $\epsilon=\epsilon(r)$, allows us to
focus refinement near the center of the grid and to reduce refinement
beyond the wave extraction zone.  We typically choose minimum and maximum 
values of $\epsilon$, for the inner and outer regions of the grid, 
respectively, and let $\log\epsilon$ vary linearly between these limits.

Unfortunately, the situation is further complicated by junk radiation in the 
initial data and time-dependent gauge effects 
as the initial data relax onto the grid.
This latter effect includes a fast moving gauge wave whose frequency
becomes higher with increasing mass ratio, $q$.
These features trigger
substantial refinement as $q$ increases.  Over-refining
on this high-frequency radiation is a waste of computational resources.
In order to limit over-refinement at early times, we have also found it 
beneficial to make $\epsilon$ 
a function of time near the beginning of the run.  

For the purposes of this work, we define three refinement functions, labelled
RF3, RF4, and RF5.  RF3 is time-dependent, spherically symmetric,
and linear in $\log\epsilon$, as shown in  Fig.~\ref{fig:rf3}.  
This refinement
function works quite well for smaller values of $q$, such as $q \lesssim 5$.
As $q$ increases, however, this 
refinement function results in prohibitively expensive runs because of
spurious waves originating around the smaller black hole. 
As a result, we introduce an additional spatial dependence to the refinement 
functions at early times, RF4 and RF5, to more sharply focus 
refinement at early times around the individual black holes.
Figs.~\ref{fig:rf4}--\ref{fig:rf5} show these two refinement functions at
$t=0$ and $t=40M$. Beyond $t=40M$, these
refinement functions become identical to RF3.
Notice that refinement is concentrated in a region around
the origin and the binary system, with $\epsilon$ increasing at 
larger radii.  Both are also 
tuned in time to allow for sufficient resolution of the outgoing radiation  
in the extraction region, while limiting refinement on the initial burst 
of spurious radiation.

We note that the definitions of these refinement functions are ad hoc, and
tuned to the specific runs reported here through experimentation.  
When used with sufficient resolution,
the refinement functions do not appear to interfere with
or change significantly the convergence properties of \dendrogr, as discussed
in Sec.~\ref{sec:tests} below, while significantly improving
the computational efficiency of the runs.  
In future work, we will explore 
generalizations that could be more widely applicable.

\subsection{Octree}

\subsubsection{Octree partitioning} 
\label{sec:octree_partition}

\begin{figure}[tbh]
\includegraphics[width=1.0\columnwidth]{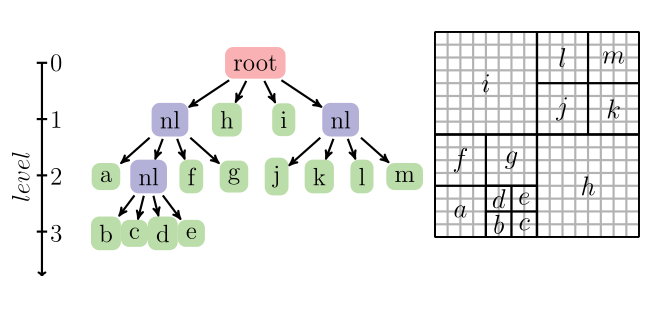}
\caption{A simple illustration of a 2D quadtree (in 3D, it would be an 
octree) as a data 
structure to represent a 2D adaptive grid. Note that we start from the 
root level, and perform a hierarchical division of each dimension 
to generate spatially varying resolution on the computational domain. In 
terms of storage, we only store the leaf nodes of the tree since non-leaf 
nodes can be computed by performing a top-down or bottom-up traversal of 
the tree. \label{fig:octree}}
\end{figure}

Octree based adaptive space discretizations (see Fig.~\ref{fig:octree}) 
are commonly used in computational science 
applications~\cite{BursteddeWilcoxGhattas11,SampathAdavaniSundarEtAl08,AhimianLashukVeerapaneniEtAl10,peanoPaper,bonsai,fernando2019massively,fernando2017machine}. Using octrees as the underlying data structure for spatial discretization is advantageous due to its simplicity, intrinsic hierarchical structure and relative ease of use in designing scalable parallel algorithms.  

In octree based adaptive multiresolution (AMR) applications, the local 
number of octants changes rapidly as the grid adapts and attempts to  
capture the spatially varying solution. This will create load imbalances 
between partitions that can reduce parallel performance. 
In order to maintain good load balancing, we need fast and efficient 
partitioning algorithms which, preferably, scale like $\mathcal{O}(n)$ 
where $n$ is the number of octants.  
Doing so will also reduce the overall communication cost between 
partitions. To this end, we use space filling curves (SFC)~\cite{sfcIntro} 
with a flexible 
partitioning scheme~\cite{fernando2017machine}. Based on the order with 
which 
these curves traverse the octants, we are able to define a partial ordering 
operator on the octree domain, which, in turn, is used to sort 
the octree.  Once this happens, higher dimensional 
partitioning reduces to a 1D problem along a curve.

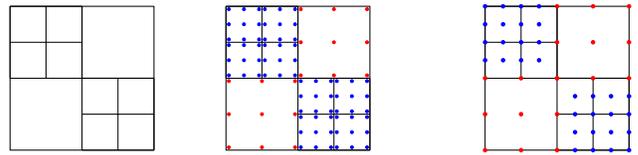
\begin{figure}[bth]
	\resizebox{0.98\columnwidth}{!}{
		\begin{tikzpicture}[scale=0.2,every node/.style={scale=0.6}]
		
		\begin{scope}[shift={(0,0)}]
		\draw[step=5] (0,0) grid +(10,10);
		\draw[step=2.5] (5,0) grid +(5,5);
		\draw[step=2.5] (0,5) grid +(5,5);
		
		\end{scope}
		
		\begin{scope}[shift={(15,0)}]
		\draw[step=5] (0,0) grid +(10,10);
		\draw[step=2.5] (5,0) grid +(5,5);
		\draw[step=2.5] (0,5) grid +(5,5);
		
		\def \r{0.1}
		\foreach \x in {0.2,2.5,4.8}{
			\foreach \y in {0.2,2.5,4.8}{
				\draw[red,fill=red] (\x,\y) circle (\r);
			}
		}
		
		\foreach \x in {5.2,7.5,9.8}{
			\foreach \y in {5.2,7.5,9.8}{
				\draw[red,fill=red] (\x,\y) circle (\r);
			}
		}	
		
		\foreach \x in {0.2,1.25,2.3,2.7,3.75,4.8}{
			\foreach \y in {5.2,6.25,7.3}{
				\draw[blue,fill=blue] (\x,\y) circle (\r);
			}
			\foreach \y in {7.7,8.75,9.8}{
				\draw[blue,fill=blue] (\x,\y) circle (\r);
			}  
		}
		
		\foreach \x in {5.2,6.25,7.3,7.7,8.75,9.8}{
			\foreach \y in {0.2,1.25,2.3}{
				\draw[blue,fill=blue] (\x,\y) circle (\r);
			}
			\foreach \y in {2.7,3.75,4.8}{
				\draw[blue,fill=blue] (\x,\y) circle (\r);
			}  
		}
		
		\end{scope}
		
		\begin{scope}[shift={(33,0)}]
		
		\draw[step=5] (0,0) grid +(10,10);
		\draw[step=2.5] (5,0) grid +(5,5);
		\draw[step=2.5] (0,5) grid +(5,5);
		
		\def \r{0.12}
		\foreach \x in {0,2.5,5}{
			\foreach \y in {0,2.5,5}{
				\draw[red,fill=red] (\x,\y) circle (\r);
			}
		}
		
		\foreach \x in {5,7.5,10}{
			\foreach \y in {5,7.5,10}{
				\draw[red,fill=red] (\x,\y) circle (\r);
			}
		}
		
		\foreach \x in {0,1.25,2.5,3.75}{
			\foreach \y in {6.25,7.5,8.75,10}{
				\draw[blue,fill=blue] (\x,\y) circle (\r);
			}
		}	
		
		\foreach \x in {6.25,7.5,8.75,10}{
			\foreach \y in {0,1.25,2.5,3.75}{
				\draw[blue,fill=blue] (\x,\y) circle (\r);
			}
		}
		
		\end{scope}	
		\end{tikzpicture}
	}
\caption{\label{fig:dg_to_cg} \small This figure shows a $2D$ example of 
the \dgn~(center) and \cgn~(right) nodal representation (with $d=2$) 
of the adaptive quadtree shown on the left. Note that the \dgn~representation has grid points that are local to each octant and contains duplicate grid points in neighboring octants. By removing duplicate and hanging grid points we get the 
\cgn~representation. In this figure, grid points are color coded based on the octant level.
}
	\vspace{-0.15in}
\end{figure}

\subsubsection{Octree construction and balancing}  
\label{sec:octree_balance}

Octree construction is the 
process of creating an adaptive octree discretization to capture a 
function $f: \Omega \rightarrow \mathcal{R}^{n}$ defined on a computational 
domain $\Omega$. The wavelet expansion of $f$ 
determines the adaptive structure for the user-specified tolerance function 
$\epsilon$. Initially, we begin from the root level of the octree and continue 
refining if the computed wavelet coefficients are greater than 
$\epsilon$. In our case, with the \BSSN\ equations, the initial grid 
is generated based 
on two puncture initial data (\S \ref{sec:id}). 
All processes begin from the root level and continue refinement until at 
least $p$ octants are produced (where $p$ denotes the number of processes). 
These $p$ octants are equally partitioned across processes. 
Further refinement occurs in an element-local fashion. As the number 
of octants increases with refinement, the octree is periodically re-partitioned
to ensure load balancing. 

We enforce an additional constraint on the octree during refinement 
which we refer to as ``2:1 grid balancing."~\cite{SundarSampathBiros08} 
This particular constraint 
enforces the condition that for a given octant in the octree, all of its 
geometric neighbors (faces, edges, and vertices) differ, at most, by a single 
level. Imposing this constraint ensures that the 
refinement structure varies smoothly through the entire grid. Moreover, we are 
guaranteed a correct interpolation stencil for points at level $j$ from 
points at level $j-1$.  As a result, this 
simplifies the subsequent mesh generation process significantly.

\subsubsection{Mesh generation} 
In order to perform numerical computations, the octree  
requires the notion of neighborhood information. 
The number of grid points placed in each  
octant depends on the degree of the finite difference stencil or polynomial 
interpolant used. For $d^{th}$-order finite differences, 
$(d+1)^3$ points are placed on each octant.  We refer to this representation 
as \dgn. 
The wavelets are calculated via interpolations of the same order.  As 
octants are shared through faces, edges, and vertices, neighboring octants 
will contain redundant informaton.  
These are efficiently identified  and then removed in order to 
get the \cgn~ respresentation (see Fig.~\ref{fig:dg_to_cg}). 
We have two mappings between these two data representations which allow for 
finite difference stencil computations of arbitrary order.  

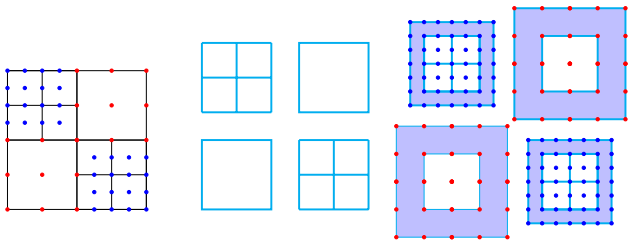
\begin{figure}[tbh]
	\resizebox{0.98\columnwidth}{!}{
	\begin{tikzpicture}[scale=0.2,every node/.style={scale=0.6}]
		
	\begin{scope}[shift={(0,0)}]
	\draw[step=5cm] (0,0) grid +(10,10);
	\draw[step=2.5cm] (5,0) grid +(5,5);
	\draw[step=2.5cm] (0,5) grid +(5,5);
	
	\def \r{0.12}
	\foreach \x in {0,2.5,5}{
	\foreach \y in {0,2.5,5}{
		\draw[red,fill=red] (\x,\y) circle (\r);
      }
	}
	
	\foreach \x in {5,7.5,10}{
	\foreach \y in {5,7.5,10}{
		\draw[red,fill=red] (\x,\y) circle (\r);
      }
	}

	\foreach \x in {0,1.25,2.5,3.75}{
	\foreach \y in {6.25,7.5,8.75,10}{
		\draw[blue,fill=blue] (\x,\y) circle (\r);
      }
	}	
	
	\foreach \x in {6.25,7.5,8.75,10}{
	\foreach \y in {0,1.25,2.5,3.75}{
		\draw[blue,fill=blue] (\x,\y) circle (\r);
      }
	}
	\end{scope}
	
	\begin{scope}[shift={(14,0)}]
	\draw[cyan,thick] (0,0) rectangle +(5,5);
	\draw[cyan,thick,xshift=2cm,yshift=2.0cm] (5,5) rectangle +(5,5);
	\draw[cyan,thick,step=2.5cm,xshift=2.0cm] (5,0) grid +(5,5);
	\draw[cyan,thick,step=2.5cm,yshift=2.0cm] (0,5) grid +(5,5);
	\end{scope}
	
	\begin{scope}[shift={(30,0)}]
	\def \r{0.12}
	\draw[cyan,fill=blue!50,fill opacity=0.5]
	(-2,-2)--(-2,6)--(6,6)--(6,-2)--cycle
	(0,0) -- (4,0)-- (4,4)--(0,4)--cycle;
	\foreach \x in {-2,0,2,2,4,6}{
	\foreach \y in {-2,0,2,2,4,6}{
		\draw[red,fill=red] (\x,\y) circle (\r);
      }
	}
	
	\draw[cyan,thick,fill=blue!50,fill opacity=0.5,even odd rule,xshift=8.5cm,yshift=8.5cm] 
	(-2,-2)--(-2,6)--(6,6)--(6,-2)--cycle
	(0,0) -- (4,0)-- (4,4)--(0,4)--cycle;
		\foreach \x in {-2,0,2,2,4,6}{
	\foreach \y in {-2,0,2,2,4,6}{
		\draw[red,fill=red,xshift=8.5cm,yshift=8.5cm] (\x,\y) circle (\r);
      }
	}

	\draw[cyan,thick,fill=blue!50,fill opacity=0.5,even odd rule,xshift=8.5cm] 
	(-1,-1)--(-1,5)--(5,5)--(5,-1)--cycle
	(0,0) -- (4,0)-- (4,4)--(0,4)--cycle;
	\draw[cyan,thick,step=2,xshift=8.5cm] (0,0) grid +(4,4);
	\foreach \x in {-1,0,1,2,3,4,5}{
	\foreach \y in {-1,0,1,2,3,4,5}{
		\draw[blue,fill=blue,xshift=8.5cm] (\x,\y) circle (\r);
      }
	}
	
	\draw[cyan,thick,fill=blue!50,fill opacity=0.5,even odd rule,yshift=8.5cm] 
	(-1,-1)--(-1,5)--(5,5)--(5,-1)--cycle
	(0,0) -- (4,0)-- (4,4)--(0,4)--cycle;
	\draw[cyan,thick,step=2,yshift=8.5cm] (0,0) grid +(4,4);	
	\foreach \x in {-1,0,1,2,3,4,5}{
	\foreach \y in {-1,0,1,2,3,4,5}{
		\draw[blue,fill=blue,yshift=8.5cm] (\x,\y) circle (\r);
      }
	}
	
	\end{scope}	
	\end{tikzpicture}
}	
\caption{\label{fig:unzip} \small This figure shows a simplified example of the octree to block decomposition and the \unzip~operation. The left figure shows the \cgn~representation.  The block decomposition is shown in the middle. Note that the given octree is decomposed into four regular blocks of different sizes. The right figure shows the decomposed blocks padded with values coming from neighboring octants with interpolated points when needed to give local uniform blocks. 
}
\vspace{-0.15in}
\end{figure}

\subsubsection{Evaluating the equations} 

\par All the field variables are defined in the compact 
octant shared, or \zipped, representation. This \zipped~representation 
allows for efficient low overhead inter-process communication. 
However, to enable finite-difference (FD) computations, it is necessary to decompose the 
adaptive octree into smaller regular grid patches or blocks. Following 
this decomposition from the octree to a block, we compute a \textit{padding} 
region for which the width depends on the maximum FD stencil radius (see 
Fig.~\ref{fig:unzip}).  
The \unzipped~representation denotes the octant local representation 
together 
with the padding region constructed from the adaptive octree. This 
\unzipped~representation is purely local to each process and discarded after 
FD stencils are evaluated (see Fig.~\ref{fig:unzip}).

\section{\label{sec:lazev} The LazEv Code}

The \lazev\ code~\cite{Zlochower:2005bj} was one of the two original 
codes to implement the moving puncture
approach~\cite{Campanelli:2005dd,Baker:2005vv}. The current version uses the 
conformal function $W=\sqrt{\chi}=\exp(-2\phi)$~\cite{Marronetti:2007wz}, 
eighth-order centered finite differencing in
space~\cite{Lousto:2007rj} and a fourth-order Runge Kutta time
integrator. 
                            
The \lazev\ code  uses the {\sc EinsteinToolkit}~\cite{Loffler:2011ay, EinsteinToolkit:2021_11}
/ {\sc Cactus}~\cite{cactus_web} /
{\sc Carpet}~\cite{Schnetter-etal-03b}
infrastructure.  The {\sc        
Carpet} mesh refinement driver provides a
``moving boxes'' style of mesh refinement. In this approach, refined
grids of fixed size are arranged about the coordinate centers of both
holes.  The {\sc Carpet} code then moves these fine grids about the
computational domain by following the trajectories of the two BHs.

The \lazev\ code implements both the BSSN~\cite{Nakamura87, Shibata95,
Baumgarte99} and CCZ4~\cite{Alic:2011gg}
evolution systems. For the tests here, we use the BSSN system.
For the gauge conditions, we use a modified 1+log lapse and a
modified Gamma-driver shift
condition~\cite{Alcubierre02a,Campanelli:2005dd,vanMeter:2006vi}, 
\begin{subequations}
\label{eq:gauge}
\begin{eqnarray}
(\partial_t - \beta^i \partial_i) \alpha &=& - 2 \alpha K,\\
\partial_t \beta^a &=& (3/4) \tilde \Gamma^a - \eta(\vec x) \beta^a \,.
\label{eq:Bdot}
\end{eqnarray}
\end{subequations}

For the function $\eta$, we choose
\begin{equation}
  \eta(\vec r) = (\eta_c - \eta_o) \exp(-(r/\eta_s)^4) + \eta_o,
\end{equation}
where
$\eta_c = 2.0/M$, $\eta_s = 40.0M$, and $\eta_o = 0.25/M$. With this
choice, $\eta$ is small in the outer zones. 
The magnitude of $\eta$ limits how large
the timestep can be with $dt_{\rm max} \propto 1/\eta$~\cite{Schnetter:2010cz}, 
Because this
limit is independent of spatial resolution, it is only significant in
the very coarse outer zones where the standard CFL
condition would otherwise lead to a large value for $dt_{\rm max}$.

We use {\sc AHFinderDirect}~\cite{Thornburg2003:AH-finding} to locate
apparent horizons.  We measure the magnitude of the horizon spin using
the {\it isolated horizon} (IH) algorithm~\cite{Dreyer02a}.
Note that once we have the horizon spin, we can calculate the horizon
mass via the Christodoulou formula
\begin{equation}
{m_H} = \sqrt{m_{\rm irr}^2 + S_H^2/(4 m_{\rm irr}^2)} \,,
\end{equation}
where $m_{\rm irr} = \sqrt{A/(16 \pi)}$, $A$ is the surface area of
the horizon, and $S_H$ is the spin angular momentum of the BH (in
units of $M^2$). 

We calculate the radiation scalar $\psi_4$ using the Antenna      
thorn~\cite{Campanelli:2005ia, Baker:2001sf}.
We then extrapolate the waveform to                             
an infinite observer location using                          
perturbative formulas from~\cite{Nakano:2015pta}.

While we use eighth-order centered difference stencils, we use a
fifth-order Kreiss-Oliger dissipation stencil and fifth-order spatial
prolongation operator (prolongation in time is second-order). We found that a rather
large dissipation coefficient of $\epsilon_{\rm dis} = 0.4$ gave the best
results.

\section{\label{sec:tests}Tests}
In this section we present some numerical results to demonstrate the
overall accuracy and performance of the \dendrogr\ framework. 
We first present results that suggest that the maximum amount
of Kreiss-Oliger dissipation should be used when solving BSSN-like
formulations of the Einstein equations.  Higher amounts of Kreiss-Oliger
dissipation increase the rate of convergence observed
in our tests.  Second, we study binary black hole mergers with mass
ratios $1 \le q \le 16$ using \dendrogr.  We show that these results
converge to equivalent solutions obtained using \lazev. 
Finally, we present results on the numerical performance of \dendrogr.
We discuss some of the refinement challenges in binary black hole
spacetimes, and show how different refinement strategies affect the overall
computational cost of the solution.

\subsection{Effects of Kreiss-Oliger Dissipation on BBH mergers}
\label{sec:KOdisstest}

\begin{table*}
\caption{Some parameters and run-time information for the runs presented in 
this paper.  All runs used wavelet tolerances $\epsilon_{\rm min} = 10^{-5}$ 
and $\epsilon_{\rm max} = 10^{-3}$. 
Runs were performed on \textsc{Expanse} at SDSC.
\label{tab:runparams}
}
\begin{ruledtabular}
\begin{tabular}{lccccccrr}
Run ID & Mass ratio & $J_{\rm max}$ & $\Delta x_{\rm min}$ & $J_{\rm max}$ & $\Delta x_{\rm min}$ & RF & SUs\footnote{Here ${\rm SU} = \sum_i c_i t_i$, where $c_i$ is the number of CPUs used for a time $t_i$, measured in hours, and $i$ is an index that runs over all of the batch jobs used to complete the run.  This measure of computational workload is not exact, as \dendrogr\ regularly rebalances the workload, which may change the number of CPUs actually used in the simulation.} & Wall Time\footnote{This is also an imperfect measure of computational performance, as the wall-clock time depends on many factors, including the number of CPU cores available for the job, and the workload per core.}\\
       & $q=m_1/m_2$ &  (BH2)  &  (BH2) & (BH1)  & (BH1) &  & (${\rm cpu}\cdot {\rm hrs}$) & (hrs) \\
\hline
q1A    &  1  & 15  & 4.069e-3 &  15 &   4.069e-3   & 2  &   --    & -- \\ 
q2RF3l &  2  & 15  & 4.069e-3 &  14 &   8.138e-3   & 3  &  5\,540  & 43 \\ 
q2RF3m &  2  & 16  & 2.034e-3 &  15 &   4.069e-3   & 3  &  41\,170  & 80 \\ 
q2RF3l &  2  & 15  & 4.069e-3 &  14 &   8.138e-3   & 4  &  5\,229  & 41 \\ 
q2RF4m &  2  & 16  & 2.034e-3 &  15 &   4.069e-3   & 4  &  39\,521  & 77\\
q4     &  4  & 16  & 2.034e-3 &  14 &   8.138e-3   & 3  &  22\,717  & 89  \\
q8RF3  &  8  & 18  & 5.086e-4 &  14 &   8.138e-3   & 3  & 485\,810  & 483\\
q8RF4  &  8  & 18  & 5.086e-4 &  14 &   8.138e-3   & 4  & 101\,915  & 318\\
q8RF5  &  8  & 18  & 5.086e-4 &  14 &   8.138e-3   & 5  &  64\,477  & 263\\
q16RF4 & 16  & 19  & 2.543e-4 &  14 &   8.138e-3   & 4  & 799\,590  & 1\,149\\
q16RF5\footnote{This job was not run to completion.} & 16  & 19  & 2.543e-4 &  14 &   8.138e-3   & 5  &  -- & -- 
\end{tabular}
\end{ruledtabular}
\end{table*}

\begin{figure}
  \includegraphics[width=\columnwidth]{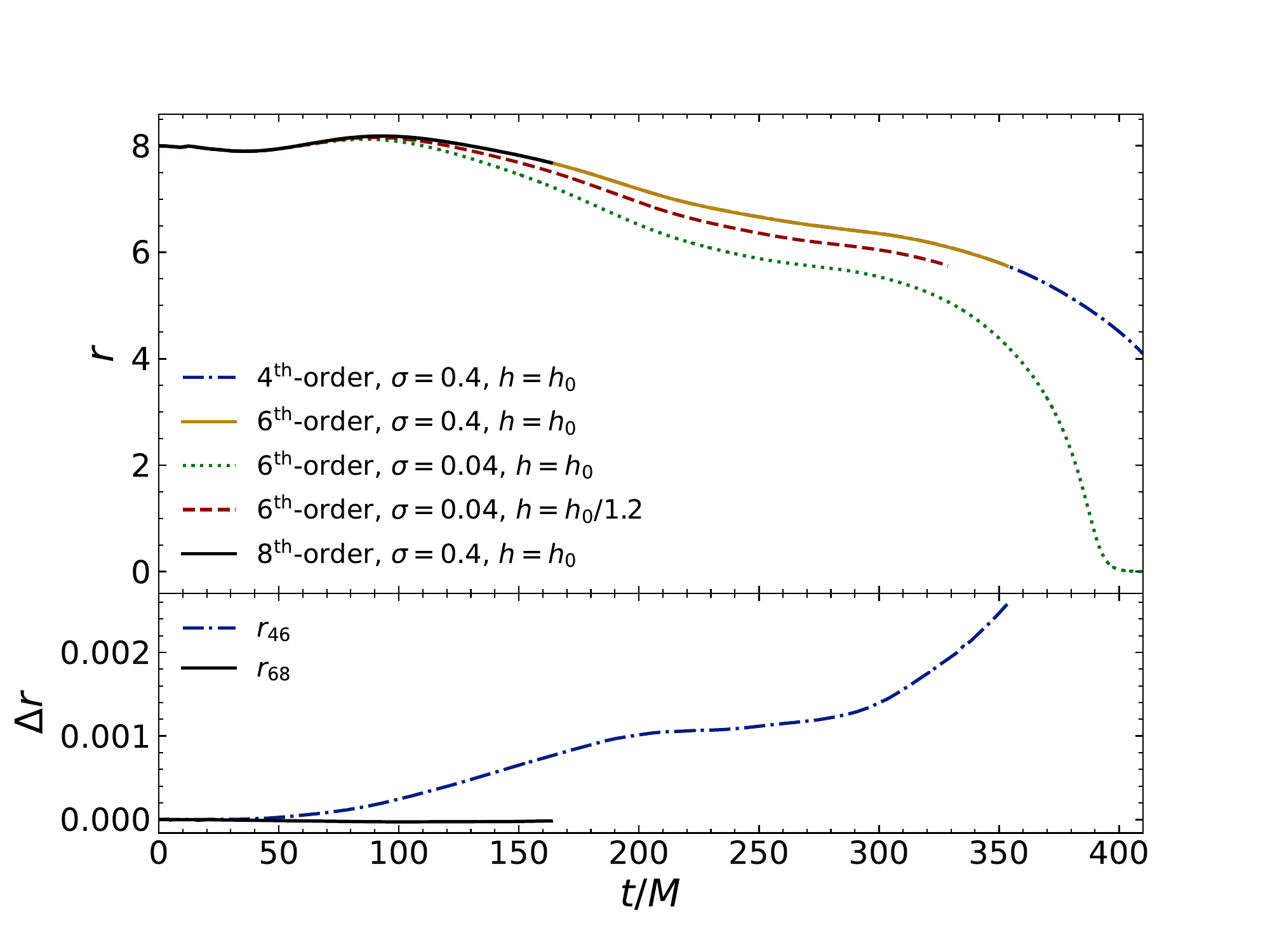}
  \caption{
  This figure shows the effect of Kreiss-Oliger dissipation on the $q=1$ binary BH 
  merger using the \lazev\ code.  Shown are results of the binary separation versus time using both different finite difference orders and different dissipation amplitudes ($\sigma$). The top frame plots the coordinate separation ($r$) between the 
  BHs as a function of coordinate time for runs with large ($\sigma=0.4$) and 
  small ($\sigma=0.04$) Kreiss-Oliger dissipation, different finite 
  difference orders ($4, 6, 8$) and at two resolutions.  The three high 
  dissipation cases computed with different FD orders are indistinguishable 
  on this plot (solid and dashed dot curves).    In contrast, the small 
  dissipation cases are significantly different, while clear convergence is seen. The lower panel plots the difference between the 4th and 6th-order results ($r_{46} = r_{\rm 4th} - r_{\rm 6th}$) and the 6th and 8th-order ($r_{68} = r_{\rm 6th} - r_{\rm 8th}$) results for $\sigma=0.4$. Note that even though the corresponding three curves in the top panel are indistinguishable, clear convergence with increasing order is seen in the lower panel. Note that $h$ indicates the coarsest resolution of the AMR grid and $h_0 = 3.3M$.
  \label{fig:r_v_diss}}
\end{figure}

Kreiss-Oliger dissipation is widely used in numerical relativity.
This dissipation is explicitly added to the numerical scheme to
eliminate high-frequency noise that can arise in the evolution,
especially near the puncture, where spacetime variables are 
non-differentiable and at refinement boundaries.  A common expectation 
is that one should minimize the amount of explicit dissipation
provided that
high-frequency noise is well controlled.  However, when performing
the initial comparisons of results from binary black hole mergers
with \dendrogr\ and \lazev, we found the opposite to be true.

In our tests, the fifth-order Kreiss-Oliger dissipation operator
in Eq.~\ref{eq:kodiss}
is added to the RHS of the semi-discrete equations
with the parameter $\sigma$, $0 \le \sigma < 1$.
We performed multiple binary black hole mergers for $q=1$ with different
values of $\sigma$ using both \lazev\ and \dendrogr.  Results from \lazev\
are shown in Fig.~\ref{fig:r_v_diss}, which plots the coordinate
separation between the two black holes.  This figure shows that the runs
with small dissipation, $\sigma = 0.04$, differ from those with large
dissipation, $\sigma=0.4$. Further, the solution with small dissipation 
converges towards those with large dissipation with increasing resolution.
Curiously, for the runs with large dissipation, the order of the
spatial finite derivatives, fourth order vs sixth, and the order of the
Kreiss-Oliger dissipation operator, fifth vs ninth, was not as important 
as the amount of dissipation, i.e., the value of $\sigma$.  Similar results 
were obtained with \dendrogr.

This result is counter-intuitive, and we are not aware of a similar discussion
in the literature.  The numerical noise in the $\sigma = 0.04$ runs was 
well-controlled, and visual inspection of the solutions did not indicate 
potential problems.  However, when solving the BSSN equations for
black hole spacetimes, better solutions at lower resolutions are obtained 
using larger amounts of explicit numerical dissipation.

\subsection{Convergence tests for Dendro-GR and LazEv} 
\label{subsec:convergence}

\begin{figure}
  \includegraphics[width=\columnwidth]{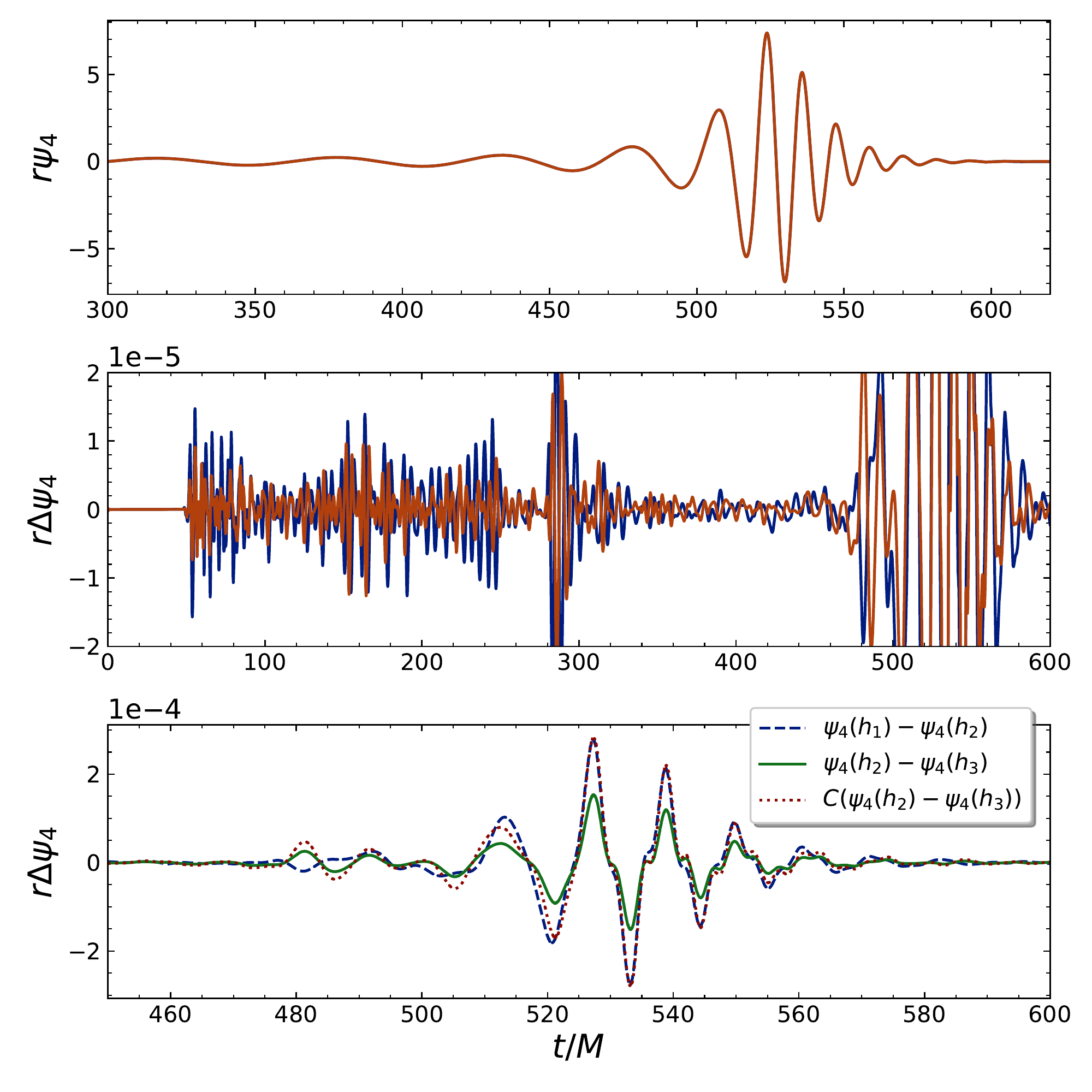}
  \caption{Convergence test of a $q=1$, non-spinning binary using 
  \lazev. The top panel shows the low and high resolution waveforms.
  The middle panel shows the differences in the waveforms between the 
  low and medium resolutions (in blue) and the medium and high resolutions 
  (in red). Because the waveforms are of comparable size, initially there 
  is small, but non-convergent noise (at these resolutions).
  The bottom panel shows the differences rescaled, assuming $3.5$
  order convergence, at the peak of the waveform. At the peak, the
  stochastic AMR noise is smaller than the truncation error. } 
\label{fig:le_conv}
\end{figure} 

Convergence is an important test not only of the computational code,
but it is also the only way to establish an estimate of the overall
error in the waveform.  To test the convergence of both codes, we
evolved initial data for an equal mass ($q=1$), non-spinning binary.
The initial data parameters are shown in Table~\ref{tab:tpid}.  
For the $q=1$ binary, we ran the \lazev\ code at three resolutions,
$\Delta x =h_0$, $h_0/1.2$ and $h_0/1.44$ with $h_0 = 3.3M$, on the
coarsest grid with nine levels of refinement.  As shown in the
center panel of Fig.~\ref{fig:le_conv}, the waveform is not initially
convergent, as relatively small stochastic errors owing to reflections
of high-frequency spurious radiation off the refinement boundaries
dominate the error.  As these high-frequency waves dissipate and
the physical signal gets larger, convergence of the error becomes
clear.  The bottom panel shows that the waveform is convergent for
the late inspiral at order 3.5.  Note that Fig.~\ref{fig:r_v_diss}
also shows convergence of the radial separation for the $q=1$ case.
The $q=2$ binaries were run with base resolutions of $h_0/1.2$ and
$h_0/1.4$, but added an additional level of refinement around the
smaller BH.  Similar convergence results were obtained for $q=2$.
Finally, for $q=4$, we ran with a base resolution of
$h_0/1.2$ and added two additional refinement levels (compared to
$q=1$) about the smaller black hole. 

\begin{figure}[tb]
\centering 
\includegraphics[width=8.5cm]{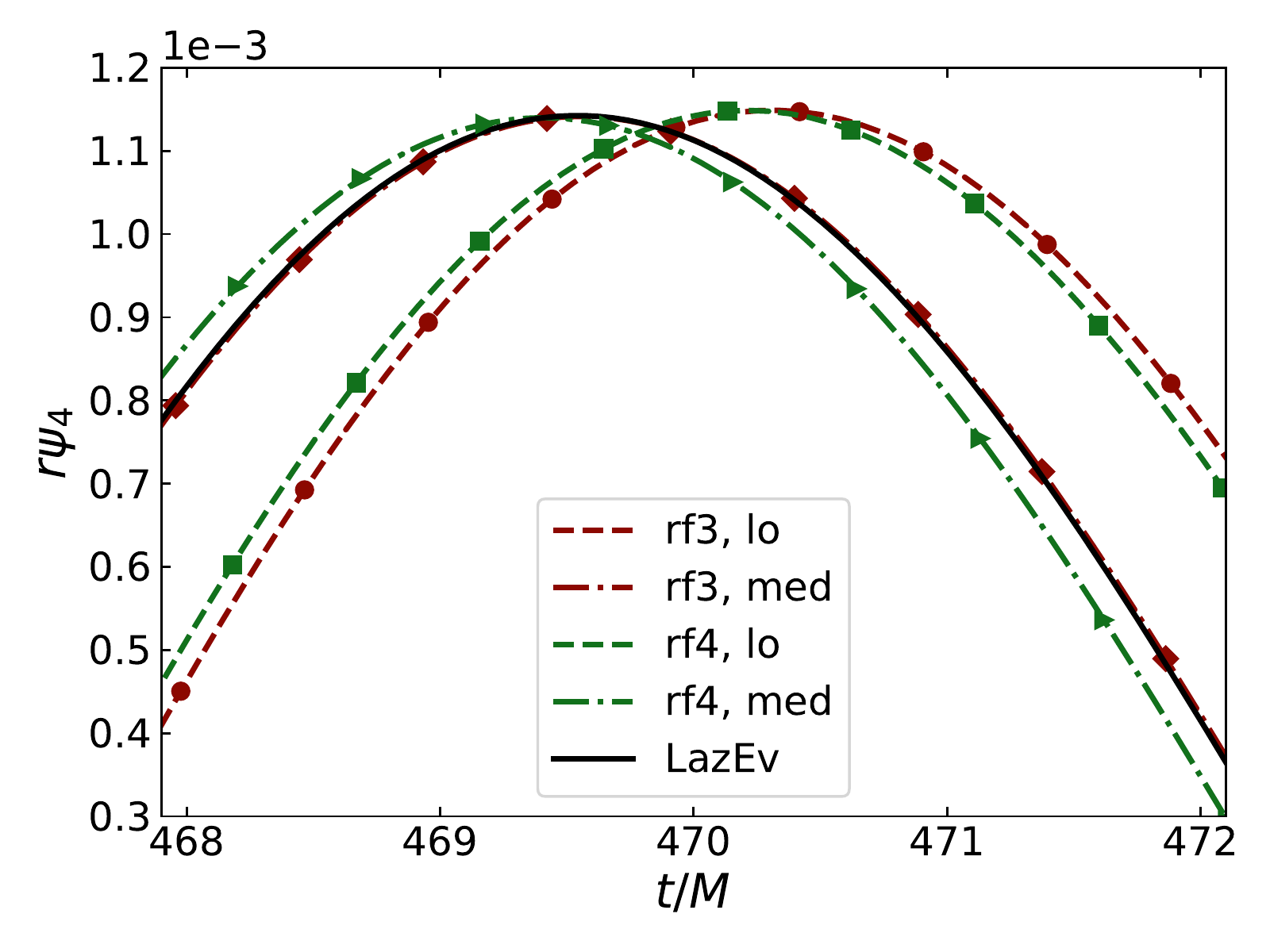}
\caption{
  This figure shows the convergence of \dendrogr\ solutions with decreasing 
  $\Delta x$.  For $q=2$, GW solutions were computed at two resolutions with 
  two refinement functions with fixed $\epsilon_{\rm min}$.  The low 
  resolution runs are plotted with dashed lines, with some representative points
  indicated with circles and squares.
  The higher resolutions runs are plotted
  with dashed dot lines, and representative points are indicated 
  with triangles and diamonds.  
  The RF3 solutions are in red and the RF4 in green.  
  Both RF3 and RF4 solutions converge to the \lazev~solution (solid black 
  line) as the maximum refinement level is increased.  The convergence is 
  largely unaffected by the choice of refinement function.  
}\label{fig:conv_de}
\end{figure}

\dendrogr\ uses an unstructured grid, and convergence is 
both more difficult to define and more challenging to demonstrate.  
Convergence depends both on the spatial resolution, $\Delta x$, as well as 
the wavelet tolerance, $\epsilon$.  Fig.~\ref{fig:conv_de} shows
the convergence of \dendrogr\ solutions (for $\psi_4$) at two resolutions,
labeled low (runs q2RF3l and q2RF4l) and medium (q2RF3m and q2RF4m) , 
for $q=2$ binaries.  The highest resolution \lazev\ $\psi_4$ is also plotted
for comparison.  With respect to changing $\Delta x$, the low and
medium resolution runs converge to the \lazev\ solution.

As mentioned above, we choose the wavelet tolerance $\epsilon$ to be a function of 
both time and space in \dendrogr.  Thus choosing different refinement functions can 
also potentially affect the solution.   Fig.~\ref{fig:conv_de} also shows
this effect by plotting results from two different wavelet refinement functions,
RF3 and RF4, for each resolution.  In this case, the effect of changing
the refinement function had a relatively small effect on the solution and
the overall runtime, see Table~\ref{tab:runparams}.

\begin{figure}[tb]
\centering 
\includegraphics[width=8.5cm]{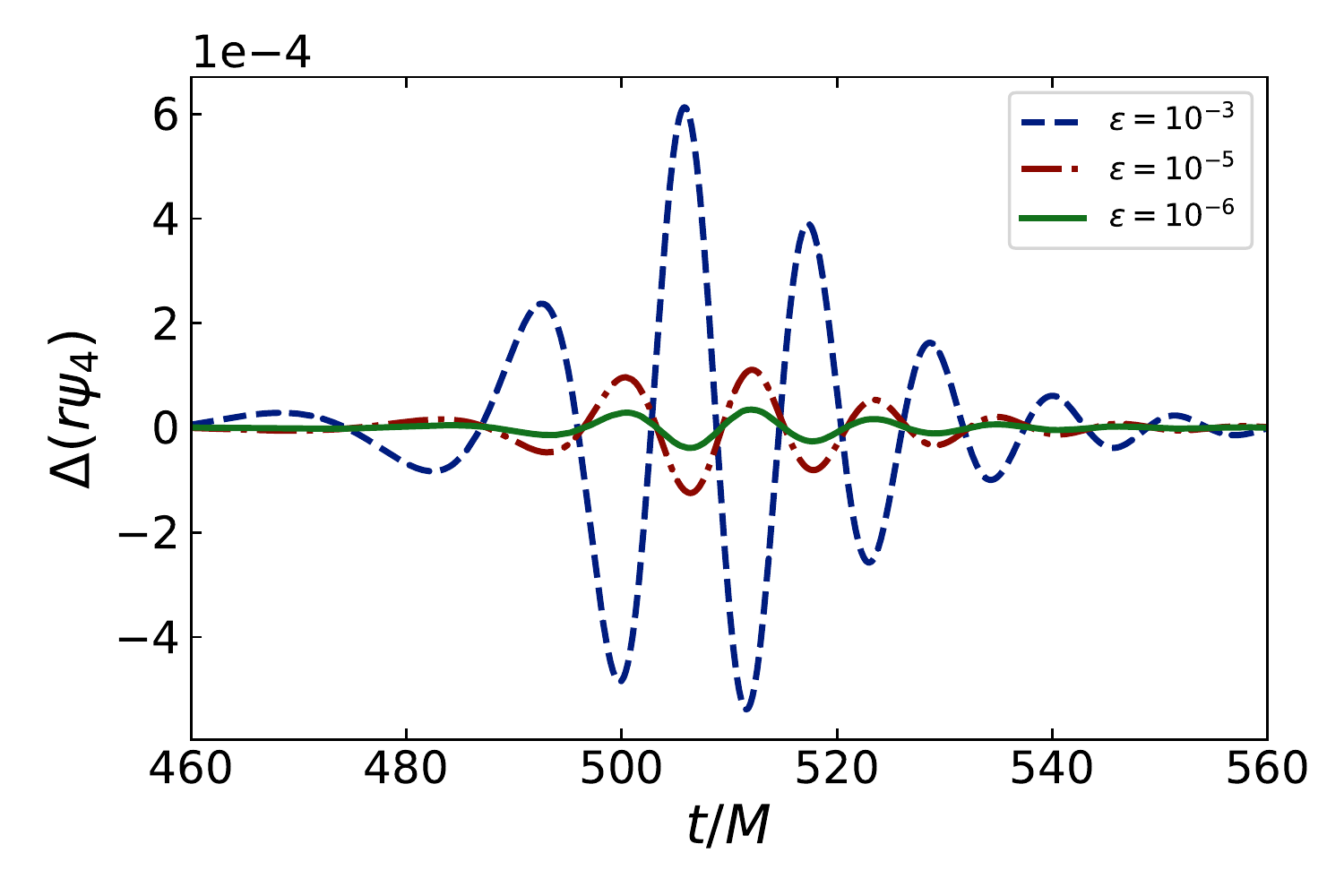}
\caption{
This figure shows convergence with respect to $\epsilon$ given a fixed 
maximum refinement
level.  The plot shows the difference between the extracted GWs using 
the \lazev\ 
and \dendrogr\ codes for the real part of the $\psi_4$ scalar
with decreasing $\epsilon$.  The \dendrogr~ solutions with decreasing $\epsilon$
converge to the \lazev~ waveforms.  Refinement function RF3 was used here.
\label{fig:q1_convergence}} 
\end{figure}

Fig.~\ref{fig:q1_convergence} illustrates the effect of only varying $\epsilon$
on the solution.  This figure compares the \dendrogr\ waveforms for
three values of $\epsilon_{\rm min} = \{10^{-3}, 10^{-5}, 10^{-6}\}$
with the highest resolution \lazev\ waveform, by plotting the difference.
Clearly the differences decrease with decreasing $\epsilon$, as smaller
values for $\epsilon$ triggers larger refined regions in the octree.
While this is a form of convergence with respect to wavelet tolerance,
the maximum refinement level, $J_{\rm max}$, and the minimum resolution, 
$\Delta x_{\rm min}$, are fixed, so this is not
convergence in the Richardson sense of the term.

\subsection{Dendro-GR binaries with different mass ratios}
\label{subsec:results}

Table~\ref{tab:runparams} gives some refinement and performance  
information for the \dendrogr\ runs reported in this paper.  The refinement
information includes the maximum allowed refinement
level, $J_{\rm max}$, the minimum resolution used in the run, $\Delta_{\rm min}$,
and the refinement function.  The performance information provides an
estimate for the total number of SUs, defined as the number of 
CPU$\cdot$hours to complete the run.  This number is approximated, because
\dendrogr\ dynamically changes the number of active threads during a run.
Finally, the table includes the total wall-clock time used to complete the
run.  While this information is valuable in providing a general view
of \dendrogr's performance, we caution that detailed conclusions cannot be
drawn.  First, the runs in this table were run over a long time period. 
During this time code changes were made, and parameters were adjusted as
we gained experience with the code.  These changes impacted the computational
costs of the runs. Secondly, wallclock times depend on the number of cores
used for each job, the final integration time, the workload per core, etc.  
For comparison, the \lazev  $q=1$ medium resolution run used 27472 SUs, 
while the high resolution run used 71651 SUs.  The \lazev\ $q=2$ medium 
resolution run used 100766 SUs, while the high resolution run used 228065 SUs.  
Finally, the \lazev $q=4$ medium resolution used 169799 SUs, while the \lazev $q=4$ high resolution used 474683 SUs. All the \lazev\ 
runs were performed on the same Intel Skylake cluster.  Note that the \lazev\ 
runs were performed at relatively high resolution to ensure that the error 
in the \lazev simulations is small compared to the \dendrogr simulations. 
These high resolution runs are required because we will use the \lazev\ 
simulations to {\it calibrate} the accuracy of the \dendrogr\ simulations.   

\begin{figure*}
    \includegraphics[width=8.5cm]{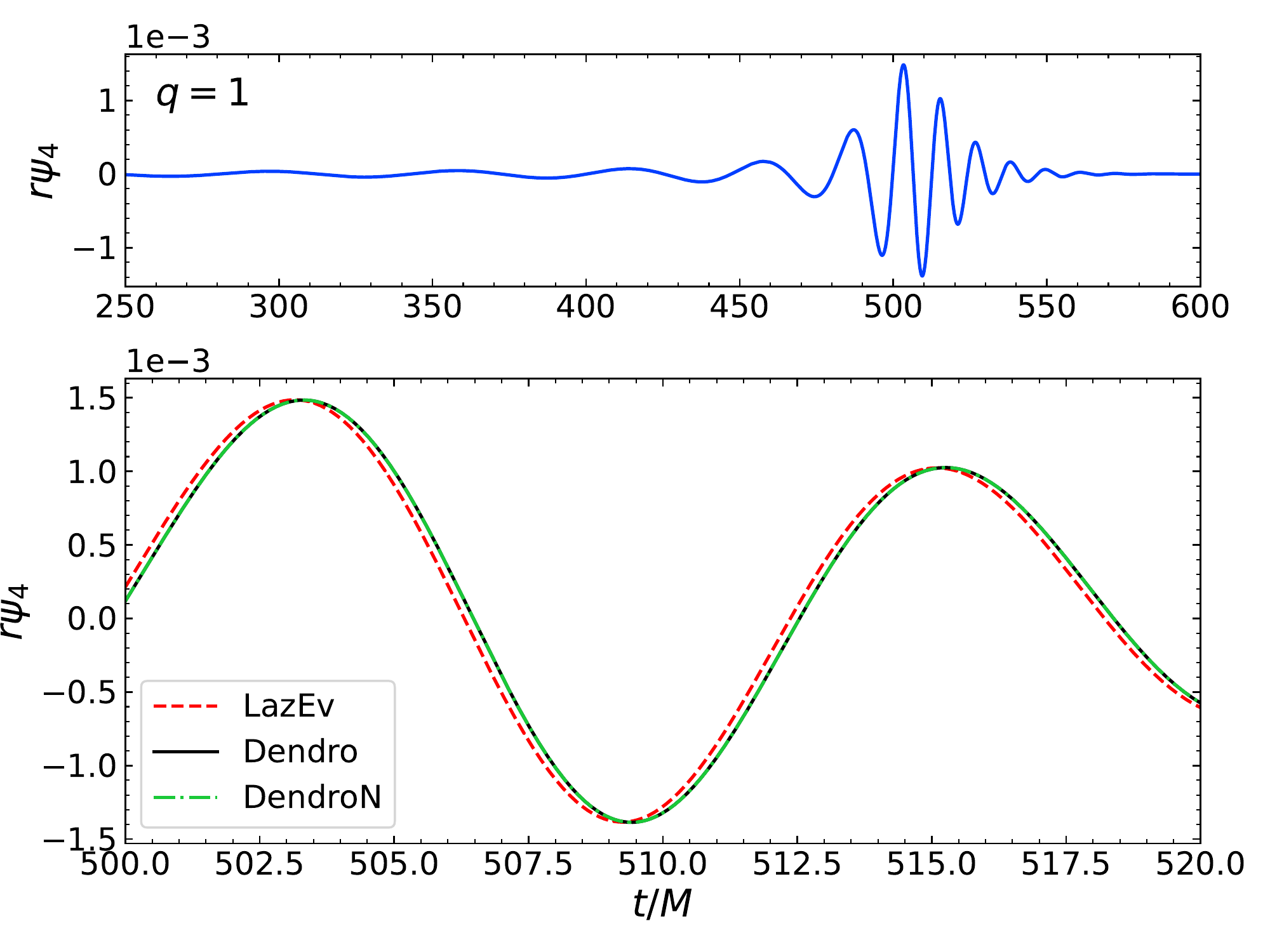}
    \includegraphics[width=8.5cm]{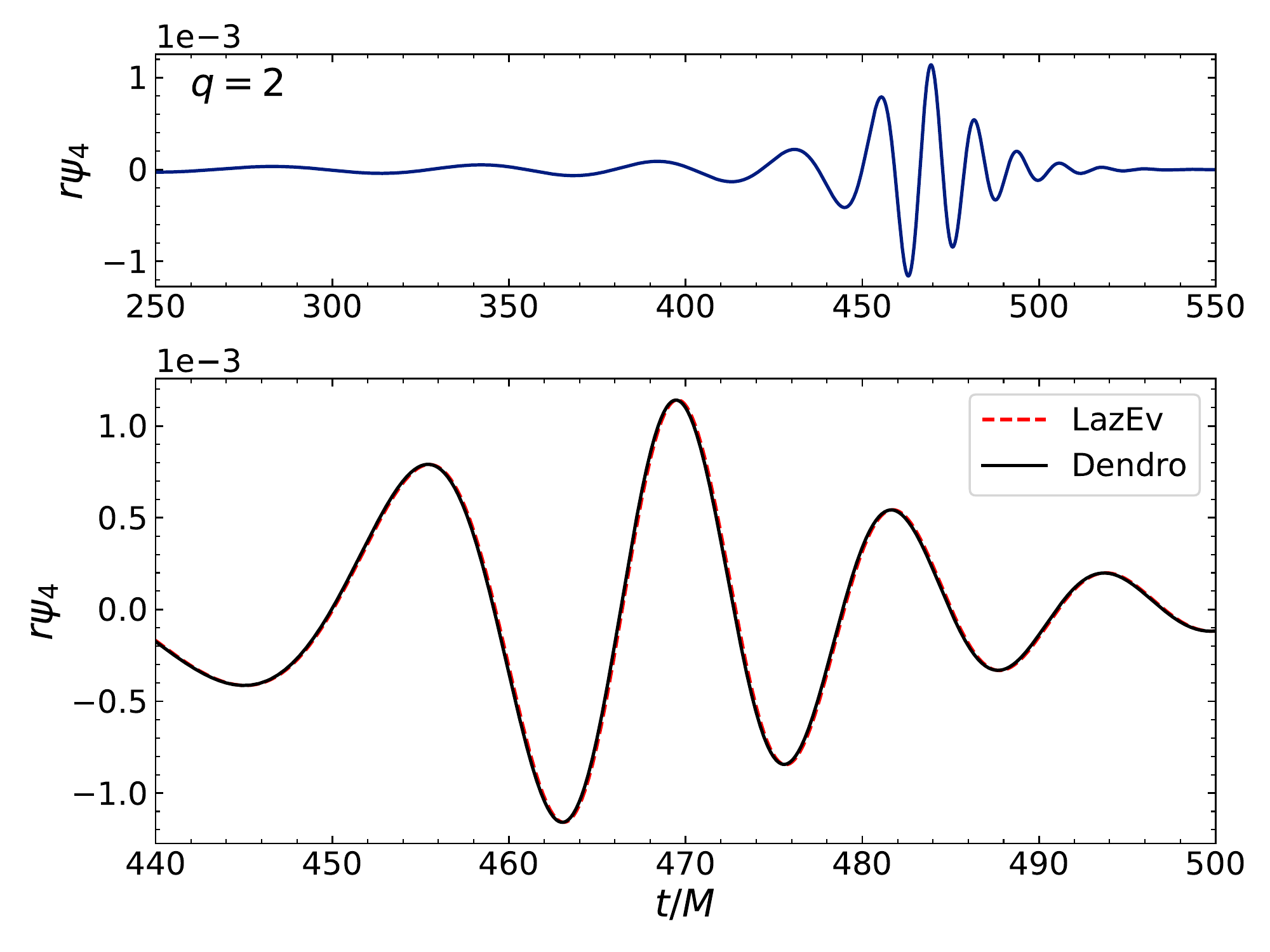}\\
    \includegraphics[width=8.5cm]{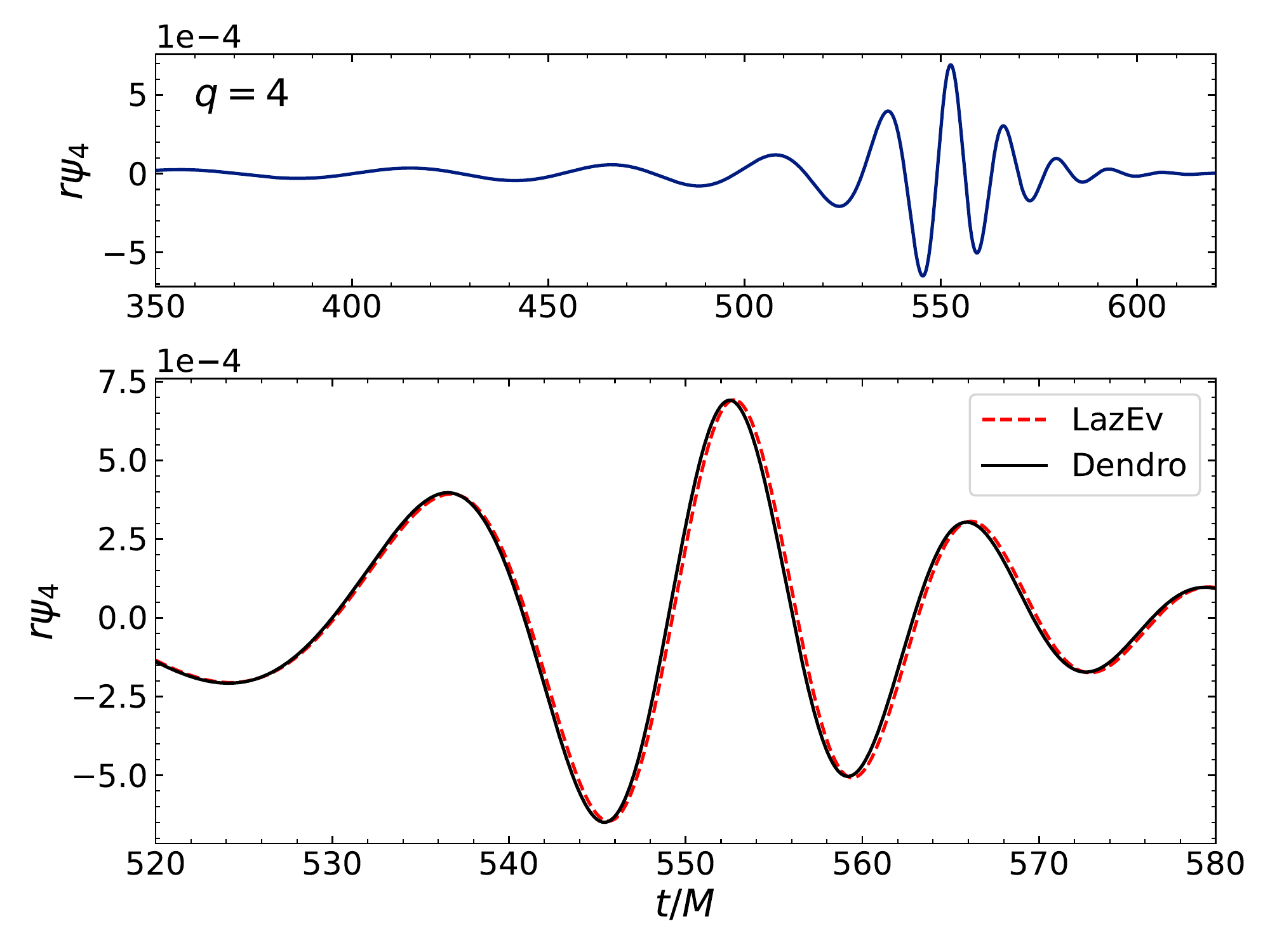}
    \includegraphics[width=8.5cm]{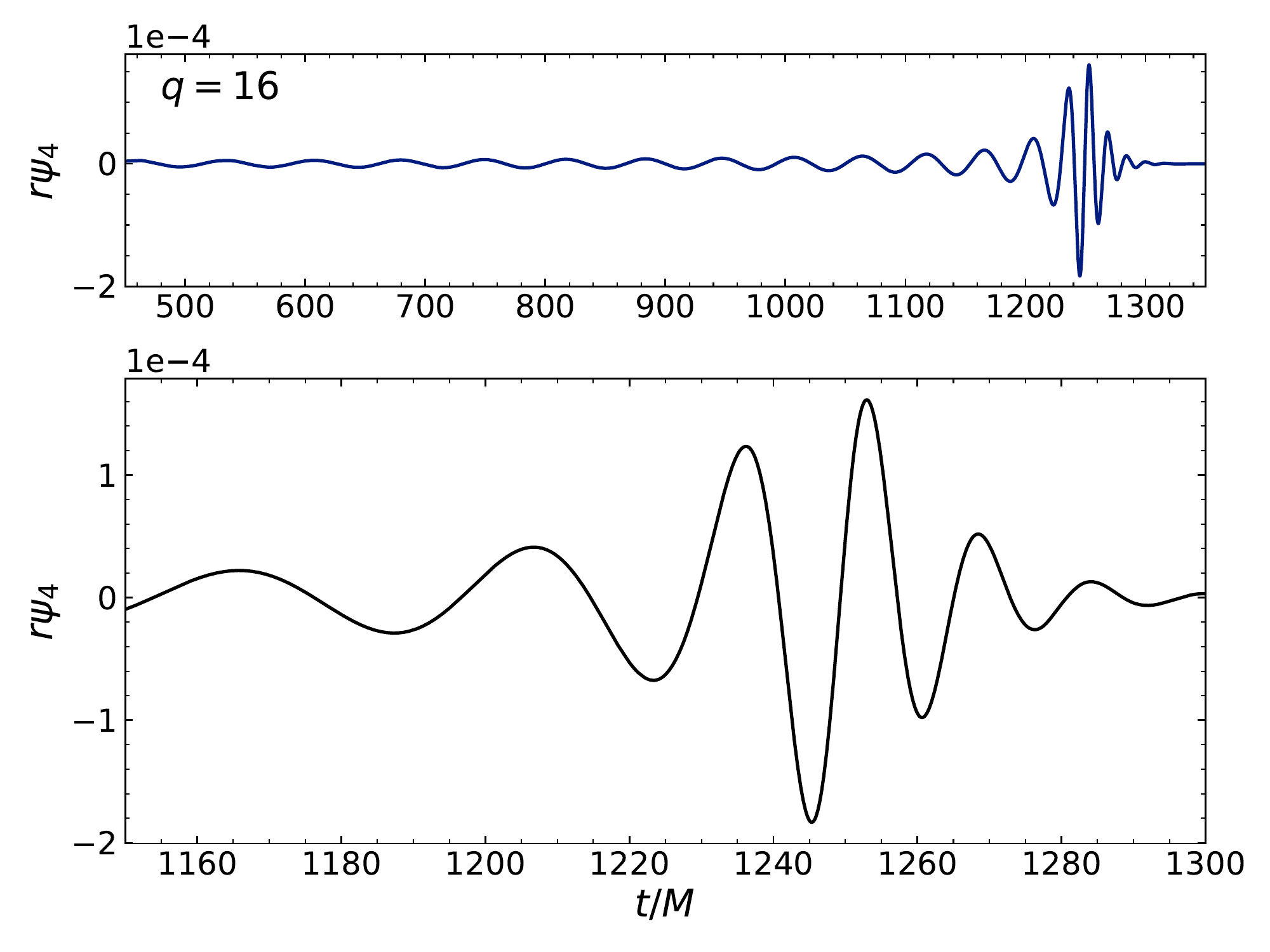}
    \caption{Plots comparing the waveform for the $q=1, 2, 4, 16$ cases. 
    \lazev and \dendrogr waveforms are given for all except $q=16$, where 
    only \dendrogr results were produced.  This figure plots the real part 
    of $\psi_4$ for the highest resolution \dendrogr and \lazev (where 
    available) runs.}\label{fig:wave_compare}
\end{figure*}

\begin{figure}
\begin{center}
    \includegraphics[width=8.5cm]{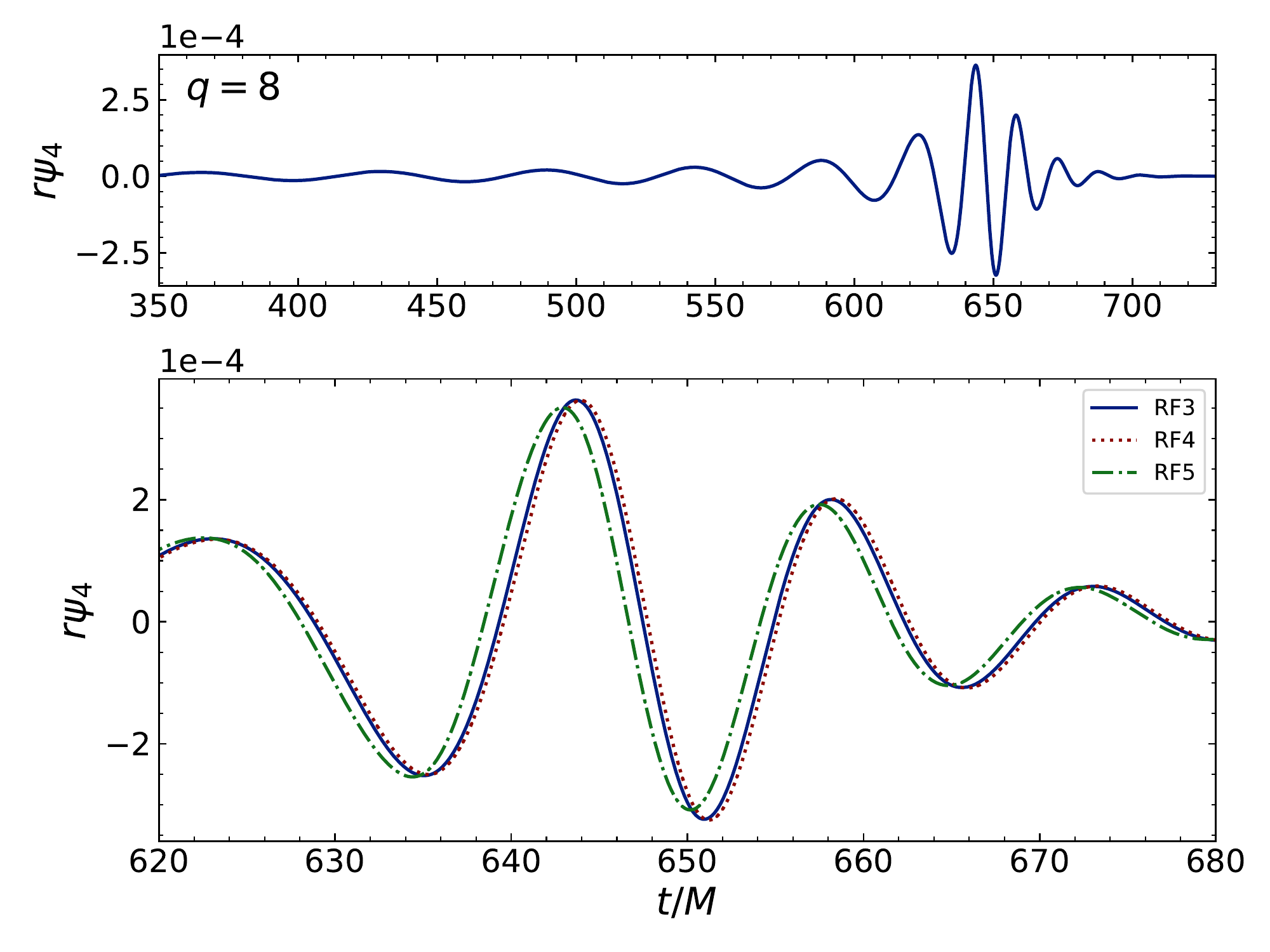}
    \caption{This figure shows the real part of $\psi_4$ for a $q=8$ black
hole binary, computed with \dendrogr\ using three different refinement 
functions with the same minimum spatial resolution.  RF3 has the 
smallest error tolerance at the center of the grid, but is very
computationally expensive.  The RF4 solution is quite similar to RF3, 
but at roughly one fifth of the computational cost (see Table~\ref{tab:runparams}).
The RF5 solution is the least expensive to compute, but for 
the parameters used here, the phase error is significant.
\label{fig:q8compare_rf}
}
\end{center}
\end{figure}

Figs.~\ref{fig:wave_compare}--\ref{fig:q8compare_rf} show gravitational
waveforms computed for non-spinning binaries with mass ratios up
to $q=16$. 
Parameters for the initial data
are shown in Table~\ref{tab:runparams}, which also gives resolution
and refinement function data, as well as the computational cost
and time to solution. 
As noted in Sec.~\ref{sec:id}, the initial data for $q\ge 2$ 
are constructed from a single family of initial data~\cite{Healy:2017zqj},
while data for $q=1$ are constructed from ad hoc parameters.
For $q=1, 2$ and $4$, the figures also show waveforms
computed with \lazev.  Due to the relativity long walltime required, 
we chose  not to complete the
correspondingly high-resolution simulations for $q=4$ simulations.
Thus the difference between the \dendrogr and \lazev waveforms for
$q=4$ may only indicate that the \lazev simulation was underresolved.
Importantly, due to its scaling, the \dendro simulations were
obtained more quickly. The binaries with $q=8$ and $q=16$ were
performed only with \dendrogr.  These figures show that \dendrogr
produces gravitational waveforms very similar to \lazev.  The mismatch
for these different waveforms are calculated below, 
in Sec.~\ref{sec:overlap}.  Unfortunately, it is difficult to draw
conclusions the accuracy of \dendrogr across different values of $q$,
because these runs were performed over a long period of time
with changing refinement strategies and a changing code base.
Many of the code changes and new approaches were motivated, in fact,
in the process of running these cases.  We were not able to go
back and rerun all cases with the same version of the code and
consistent refinement criteria.

As discussed in Sec.~\ref{sec:rf}, a gauge wave propogates
across the computational domain at early times,
as the coordinates transition from the Bowen-York gauge conditions,
used to calculate the initial data~\cite{Ansorg:2004ds}, 
to the puncture gauge conditions used in the evolution.
The wavelength of the gauge wave is related to the black hole size,
and thus the frequency of this unphysical wave increases with mass
ratio, $q$.
The high frequency wave triggered excessive refinement
at the beginning of the higher mass ratio runs, particularly for 
$q\ge 8$, prompting our experimentation with
different wavelet refinement functions.  We ran simulations of
the $q=8$ binary with three different refinement functions, and plot
the resulting waveforms in Fig.~\ref{fig:q8compare_rf}. RF3 uses
$\epsilon_{\rm min}$ over a larger volume of the grid, while RF4
and RF5 allow for a larger wavelet tolerance over a larger region
of the grid.  Consequently, RF3 likely gives a more precise
solution but at a greater computational cost as it may tend to 
over-refinement.  While RF4 and RF5
are more computationally efficient, differences in the 
waveforms become noticable.  For $q=16$, RF3 was too expensive,
and this run was only done with RF4 and RF5.  The RF5 run was
not completed, as the differences in results for RF3 and
RF5 for $q=8$ were large.
Interestingly, the differences in RF3 and RF4 occur only at the
initial time.  After $t=40M$, both refinement functions are identical.
The phase differences seen in the figure seem to arise solely from small 
variations in the refinement at early times.  

\begin{figure}
\begin{center}
    \includegraphics[width=8.5cm]{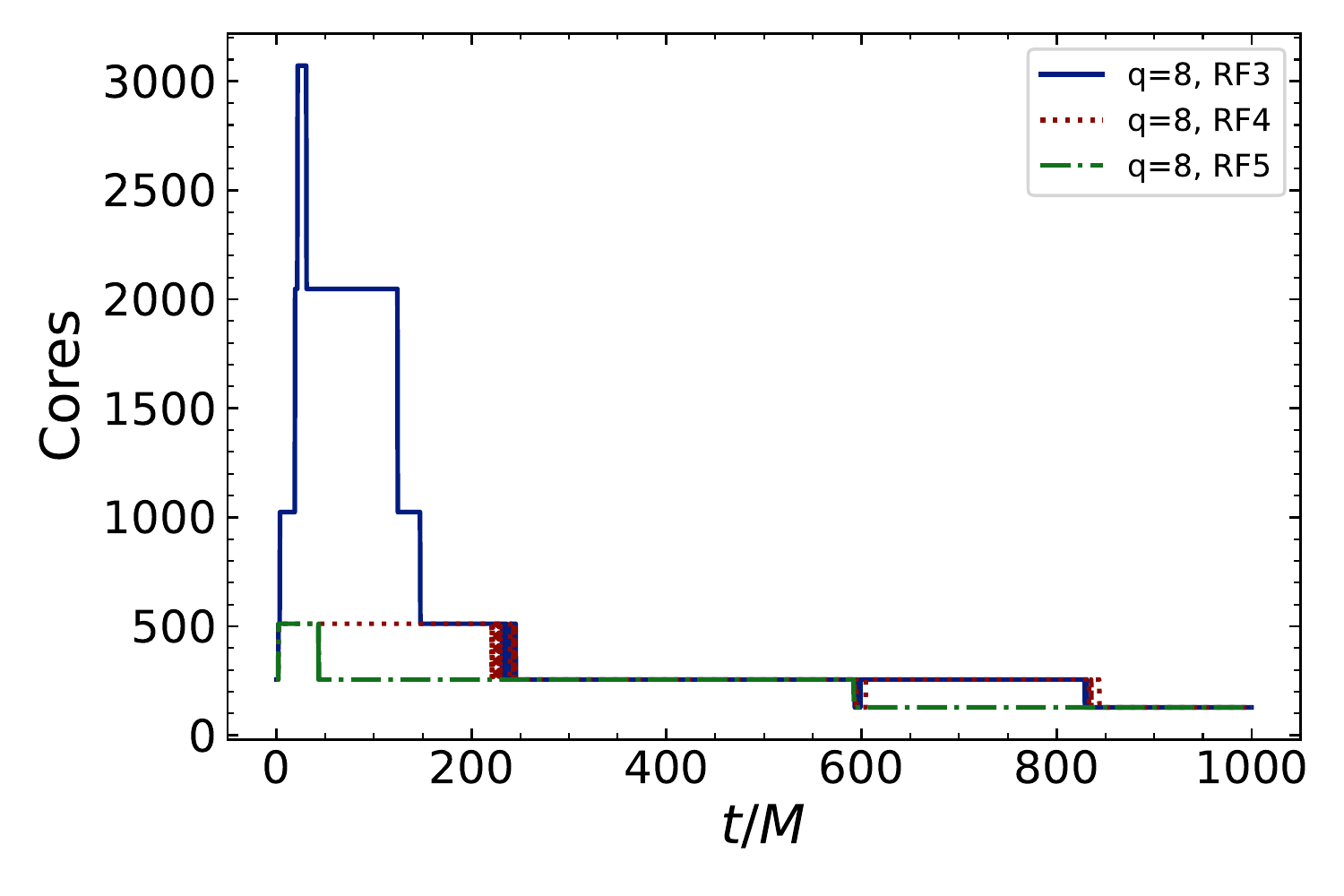}
    \caption{
This figure shows the effect of different refinement functions
on the computational cost of evolving a $q=8$ black hole binary 
with \dendrogr\ by plotting  the number of
computational cores used during the simulation as a function of the 
computational time.  The three refinement functions, RF3, RF4, and
RF5, differ only for $t \le 40~M$.  Thus, the differences arise
primarily in how the initial junk radiation and gauge waves are
resolved. As shown in Fig.~\ref{fig:q8compare_rf}, 
the gravitational wave results from RF3 and RF4 are similar, 
although the maximum workload for RF4 was about 6 times smaller 
than the maximum workload for RF3. The results for RF5, while the
most efficient run, show larger differences from the other two cases.
    \label{fig:q8cores}}
\end{center}
\end{figure}

\begin{figure}
\begin{center}
    \includegraphics[width=8.5cm]{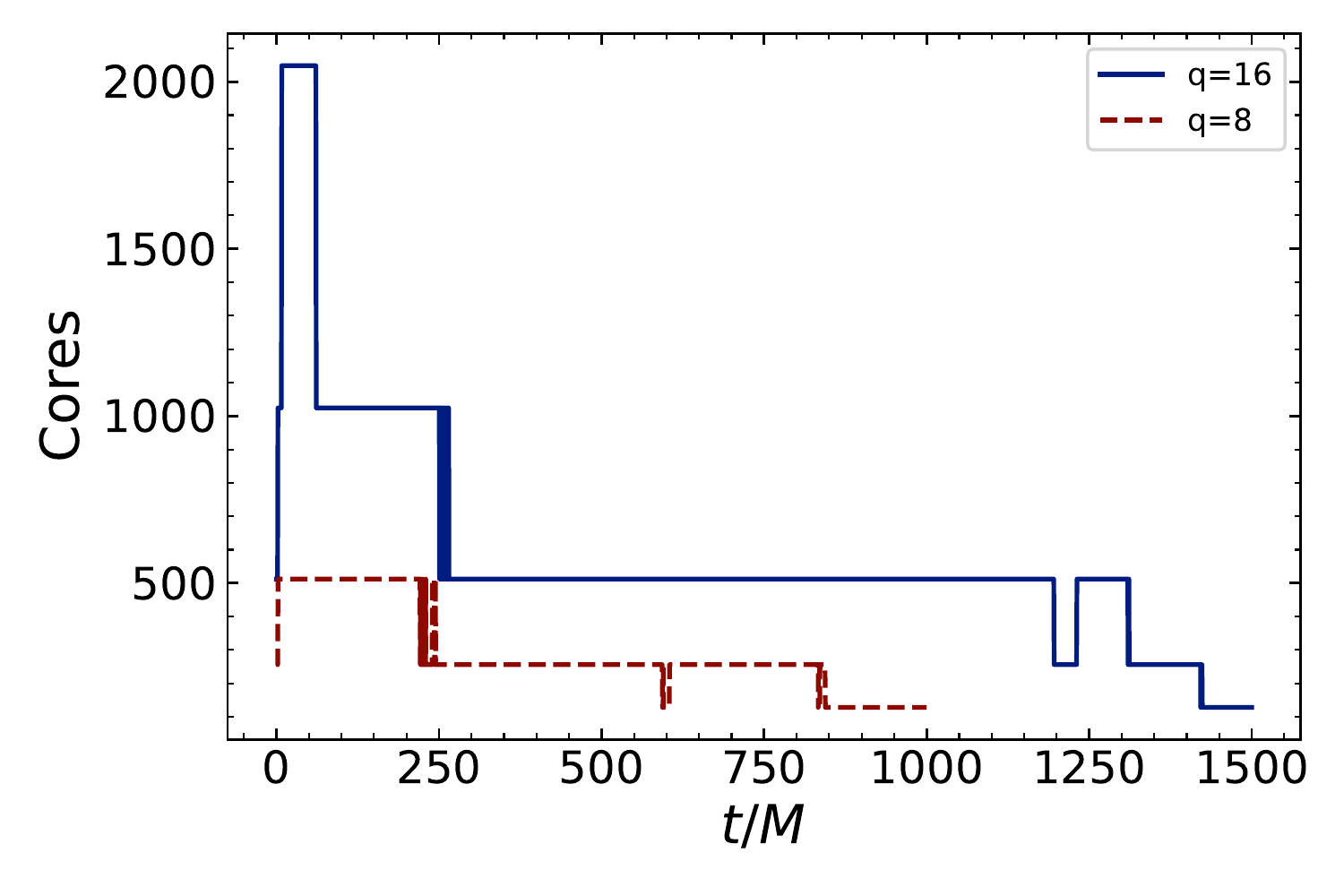}
    \caption{This figure shows the number of computational cores used in 
$q=8$ and $q=16$  binary black holes merger simulations with RF4.  
    \label{fig:cores}}
\end{center}
\end{figure}

The number of computational cores used in the $q=8$ and $q=16$ runs
are plotted in Figs.~\ref{fig:q8cores} and \ref{fig:cores}, respectively.
\dendrogr\ regularly repartitions the computational workload across
the available cores.  To balance the communication cost between cores, 
it will use fewer cores than the total number 
available if the workload per core drops below some threshold.
In these runs, over-refinement on the high frequency gauge wave and junk
radiation remains a problem, and causes the large increase in demand for
computational resources at the beginning of the run.  As this radiation
moves beyond gravitational wave extraction region, the grid is coarsened
and the runs become much more efficient.

\subsection{Overlaps}
\label{sec:overlap}

When using numerical waveforms for gravitational wave data analysis, 
numerical convergence provides an important estimation of the error in the
numerical solution.  Ideally, the convergence error, determined by comparing 
the solutions computed at two different resolutions, is smaller than 
the other errors in the analysis.  However, convergence testing overlooks
the frequency response of a real-world detector.  The overlap provides
a way to compare two waveforms as measured in a detector with a given 
frequency response.  In addition, the overlap helps us to determine the
computational resources required to simulate a particular configuration.
This allows us to determine how similar two waveforms, computed at 
different resolutions or with different codes, are to one another.

For this analysis, we will measure the consistency of two waveforms using the
\texttt{CreateCompatibleComplexOverlap} function in {\sc LaLSimUtils}
(which is freely available)~\cite{Pankow:2015cra, Lange:2018pyp}.  This function
automatically optimizes over both time translations and phase shifts.
Because of this, the mode-by-mode mismatch allows for the phase
shifts of different modes to be inconsistent. That is, one expects
each $m$-mode to be shifted by $m\phi$.

Internally, this function uses the inner product                            
\begin{equation}                                                              
  \left \langle h_1| h_2 \right \rangle = 2 \,  \int_{-\infty}^{\infty} \frac{h_1^*(f) h_2(f)}{S_n(f)}  \label{eq:overlap}
\mathrm{d}f,
\end{equation}
where  $h(f)$ is the Fourier transform of the complex waveform $h(t)$ and we use the Advanced-LIGO
design sensitivity \emph{Zero-Detuned-HighP} noise 
curve~\cite{ligonoisecurve} with $f_{\text{min}} = 20$~Hz and  
$f_{\text{max}} = 2000$~Hz. This inner product  is then
further maximized
over time and phase shifts as described in~\cite{Ohme:2011zm}
\begin{equation}              
   \left \langle h_1| h_2 \right \rangle = \max_{t_0,\phi_0}\left[\, 2\,\left|  \int_{-           
\infty}^{\infty} \frac{h_1^*(f) h_2(f)}{S_n(f)} \mathrm{d}f \right|\,  \right].
\end{equation}            
The overlap of two waveforms is then given by
\begin{equation}            
  \mathcal{O}=    \frac{   \left \langle h_1| h_2 \right \rangle }{\sqrt{   \left \langle h_1| h_1       
\right \rangle  \left \langle h_2| h_2 \right \rangle  }} 
\end{equation}                     
and the mismatch is  given by
\begin{equation} \label{eq:mm_int}
  \mathcal{M} = 1 - \mathcal{O}.
  \end{equation}

Because Eq.~(\ref{eq:overlap}) directly involves the detector's
noise sensitivity curve, $S_n(f)$, the mismatch is a function of the actual
frequency waveform and is not invariant under a change in the total
mass of the system. Depending on the total mass, only a portion of the 
waveform may be in the detector's sensitivity region.  As an example,
for systems with small total mass, LIGO is most sensitive to the low-frequency
portion of the signal at early times.  For systems with a large total
mass, however, LIGO is most sensitive to the high-frequency merger and early 
ringdown portion of the waveforms. 

\begin{figure}[tb]
\centering
\includegraphics[width=8.5cm]{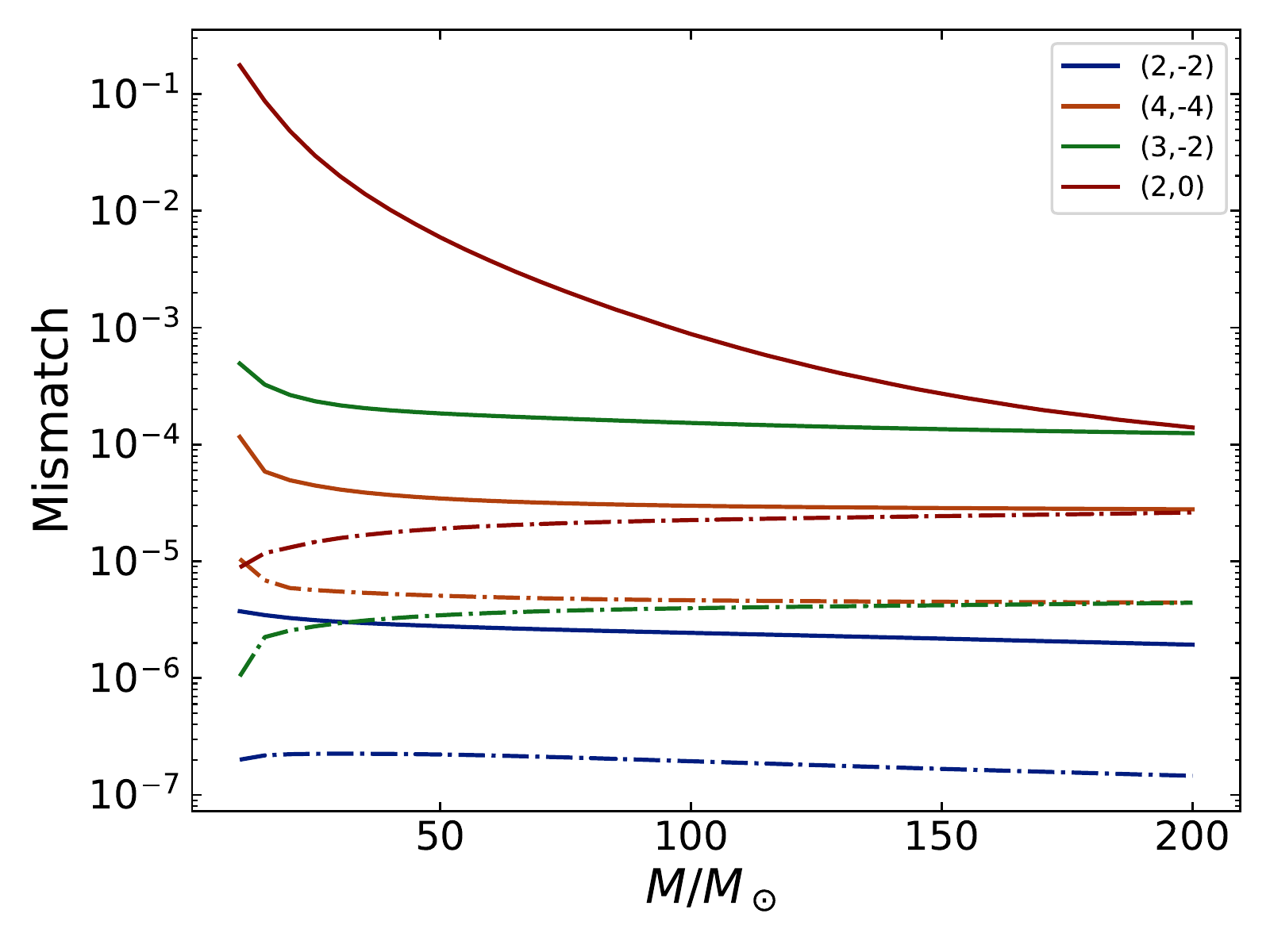}
\caption{
This figure shows the overlaps in gravitational waveforms
for various modes of the \dendrogr\ and
  \lazev\ solutions for a $q=1$ binary.
The dot-dashed lines show the overlaps of the \lazev\
medium resolution waveforms against the \lazev\ high-resolution waveforms.
The solid lines show the overlaps of the \dendrogr\ waveforms against the
  \lazev\ high-resolution waveforms. The modes are presented in order of decreasing amplitude. The $(2,0)$ mode, which fails our accuracy goal of ${\cal M} < 0.005$ is subdominant to all the other modes.
\label{fig:q1_overlap}}
\end{figure}

\begin{figure}[tb]
\centering
\includegraphics[width=8.5cm]{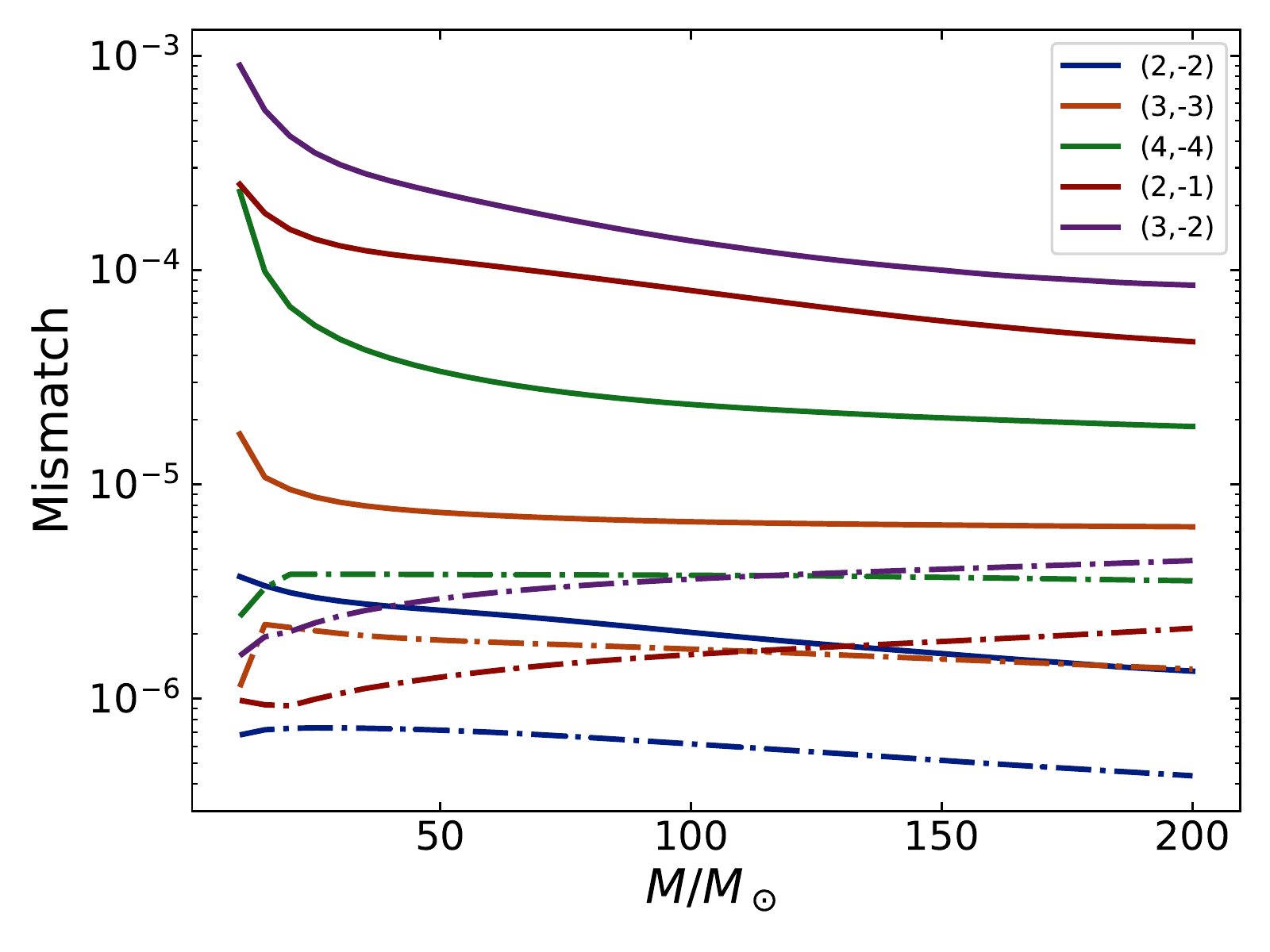}
\caption{
This figure shows the overlaps in gravitational waveforms
for various modes of the \dendrogr\ and
\lazev\ solutions for a $q=2$ binary.
The dot-dashed lines show the overlaps of the \lazev\
medium resolution waveforms against the \lazev\ high-resolution waveforms.
The solid lines show the overlaps of the \dendrogr\ waveforms against the
  \lazev\ high-resolution waveforms. The modes are presented in order of decreasing amplitude (i.e., (3,-3) is subdominant to (2,-2), (4,-4) to (3,3), etc.).
\label{fig:q2_overlap}}
\end{figure}

\begin{figure}[tb]
\centering
\includegraphics[width=8.5cm]{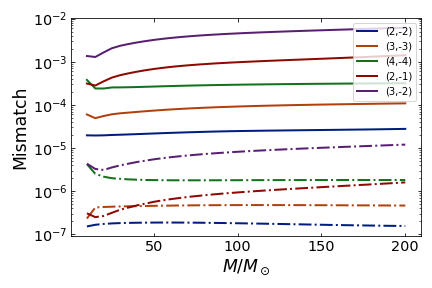}
\caption{
This figure shows the overlaps in gravitational waveforms
for various modes of the \dendrogr\ and
\lazev\ solutions for a $q=4$ binary.
The dot-dashed lines show the overlaps of the \lazev\
medium resolution waveforms against the \lazev\ high-resolution waveforms.
The solid lines show the overlaps of the \dendrogr\ waveforms against the
  \lazev\ high-resolution waveforms. The modes are presented in order of decreasing amplitude (i.e., (3,-3) is subdominant to (2,-2), (4,-4) to (3,3), etc.).
\label{fig:q4_overlap}}
\end{figure}

We use the mismatch, $\cal M$, to  compare the \dendrogr\ and \lazev\ 
waveforms. A mismatch of ${\cal M} < 0.005$ was determined in 
Ref.~\cite{Lindblom:2008cm} to be minimally acceptable for advanced LIGO 
analysis. Ideally, a mismatch ${\cal M} \ll 0.005$ is desired. However, this 
limit of $<0.005$ is for the net mismatch in the observed waveforms (i.e., 
after summing all modes). Setting the mismatch tolerance to $<0.005$ 
for all subdominant modes is therefore more restrictive than required. Here, 
we want to use the mismatch between the \lazev and \dendrogr simulations to 
measure the truncation error in the \dendrogr simulations. This is only true 
if the error in the \lazev simulations is much smaller than the \dendrogr 
simulations. To an attempt to guarantee this, we require that the corresponding 
\lazev-to-\lazev mismatches (between the medium and high resolutions) are 
much smaller than the corresponding \lazev-to-\dendrogr mismatches
(we note that a small \lazev-to-\lazev mismatch
may not account for all possible global errors).
When 
this is not the case, the mismatch between the two codes is not a measure 
of the error of the \dendrogr simulations.
 Figs.~\ref{fig:q1_overlap} - \ref{fig:q4_overlap} show
the overlaps for different modes of $\psi_4$ computed with \dendrogr\
and \lazev, for the $q=1$, $q=2$, and $q=4$ binaries, respectively. 
In particular, we use the high-resolution \lazev\ solutions
as the base solutions for comparison with the \dendrogr\ and medium-resolution
\lazev\ waveforms. The figure shows the modes in order of decreasing amplitude. The $q=1$ subdominant \dendro\ $(2,0)$ mode shows a significant mismatch with the corresponding \lazev mode. All modes with amplitude larger than the $(2,0)$ mode show much smaller mismatches between \dendro and \lazev. The $q=2$ and $q=4$ comparisons show similar behavior (here, more modes are non-trivial, and the $(2,0)$ mode is subdominant to all modes shown).

\begin{figure}[tb]
\centering
\includegraphics[width=8.5cm]{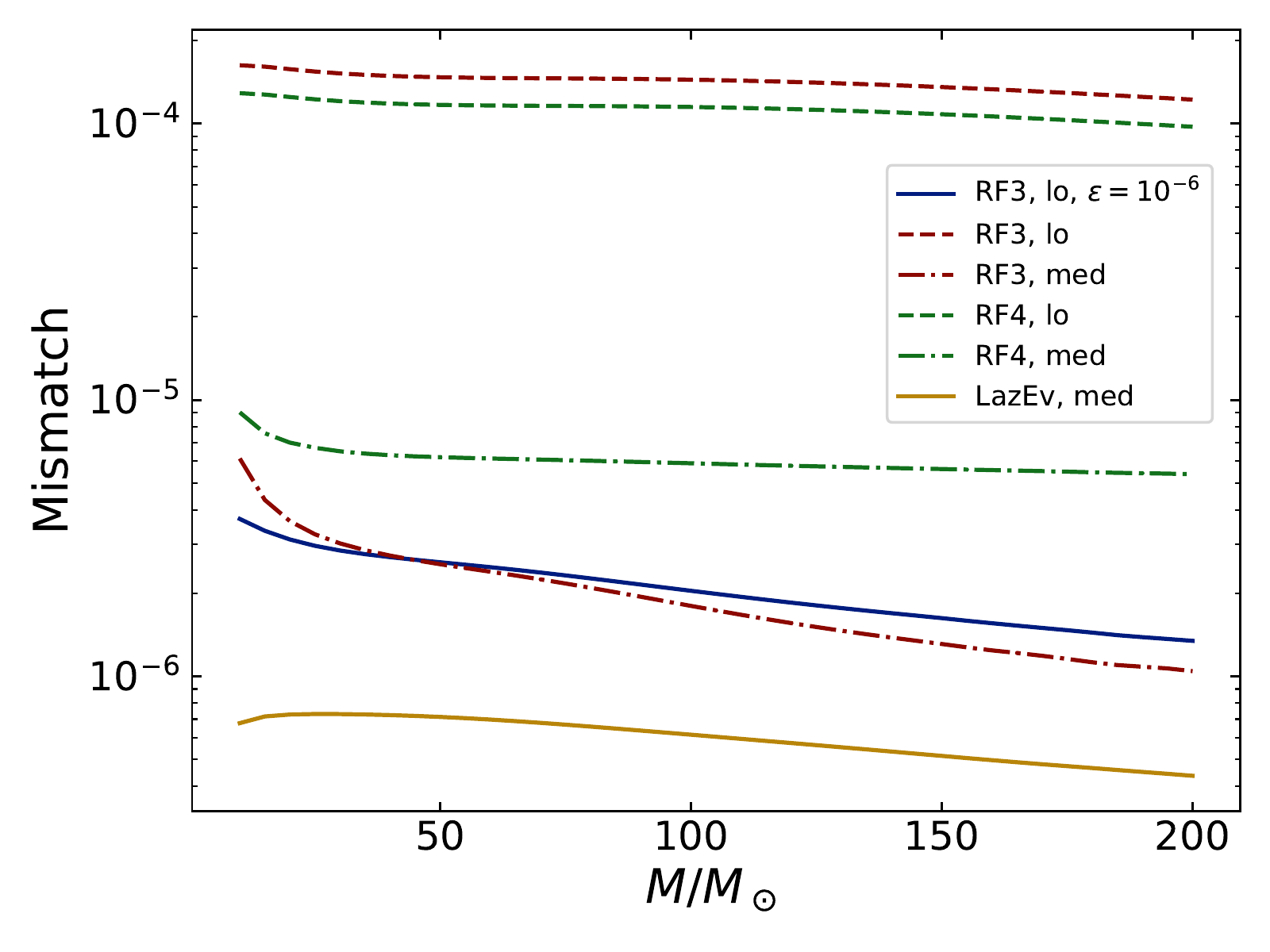}
\caption{
This figure shows the overlaps in the $(\ell,m) = (2,-2)$ mode of $\psi_4$
computed with various \dendrogr\ refinement criteria against the 
high-resolution \lazev\ solution for a $q=2$ binary.
The blue line is computed with $\epsilon = 10^{-6},$ and all other cases
use $\epsilon = 10^{-5}$.  As expected, the waveforms computed with the
more expensive RF3 refinement function (red) have better overlaps to \lazev
than those computed with RF4 (green).
Finally, the tan solid line shows the overlap of the \lazev\ 
medium and high resolution waveforms. 
\label{fig:q2_overlap_runs}}
\end{figure}

\begin{figure}[tb]
\centering
\includegraphics[width=8.5cm]{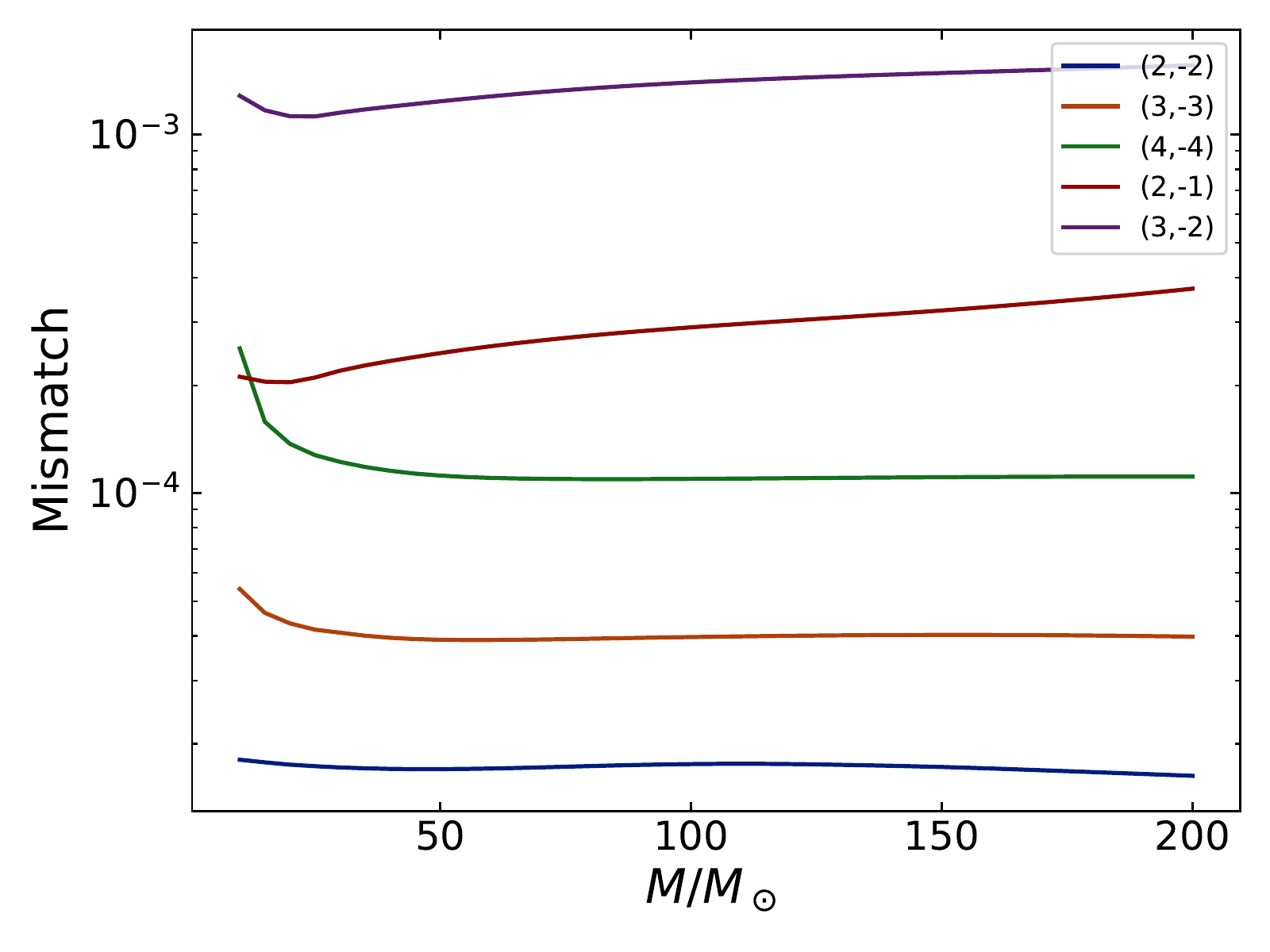}
\caption{
This figure shows the overlaps in the \dendrogr\ 
$q=8$ gravitational waveforms computed with the RF3 and RF4 refinement functions.
The waveforms are presented in order of decreasing amplitude.
As expected, the dominant modes have smaller mismatches than the subdominant modes.
\label{fig:q8_overlap}}
\end{figure}

Fig.~\ref{fig:q2_overlap_runs} shows the overlaps
in the $(\ell,m)=(2,-2)$ mode of $\psi_4$ for 
\dendrogr\ waveforms computed with different refinement criteria.
These overlaps compare solutions computed with different refinement functions,
RF3 and RF4, with a refinement tolerance $\epsilon=10^{-5}$.  The figure 
also shows the overlap for a solution computed with RF3 and the refinement 
tolerance set to $\epsilon=10^{-6}$. The main result from this figure is that higher resolution (more restrictive error tolerance) leads to a better agreement between \lazev and \dendro. Finally, the overlap of the 
medium-resolution \lazev\ solution is shown.  Consistent with the earlier
convergence results, the \dendrogr\ runs match the high-resolution
\lazev\ solution well, and the RF3 solution is slightly closer.  Finally,
Fig.~\ref{fig:q8_overlap} shows the overlap of the 
\dendrogr\ solutions for the $q=8$
binary computed with the RF3 and RF4 refinement functions.

\section{\label{sec:discuss}Discussion}
This paper presents binary black hole evolutions performed with 
\dendrogr\ for different mass ratios up to $q=16$.  We present validation
tests in comparison with results from \lazev, and we give performance 
information for these runs.

While the focus of this paper is on evolving binary black holes with \dendrogr,
the first result that we presented has general applicability in the numerical
relativity community.  We found that in binary black hole evolutions with the
\BSSN\ formalism, the rate of convergence is increased when a large
amount of Kreiss-Oliger dissipation is added to the solution.  In our tests,
runs with a dissipation parameter of $\sigma = 0.4$ had a better rate of
convergence than runs with $\sigma = 0.04$, where $\sigma$ is bounded 
by $\sigma < 1$ for numerical stability.

While the performance and scaling of \dendrogr\ is very good, we are 
currently working on additional improvements.  In particular, we are 
improving the unzip process, in which an octant of the tree in locally
expanded to a uniform Cartesian grid, to reduce the communication overhead.
As shown in this paper, we have started exploring different ways to
control the refinement algorithm, especially during the initial times of
an evolution.   We want to
improve the computational performance of \dendrogr, while not sacrificing
the accuracy of the solutions.  While we have had some initial success, much
more work remains to be done.  We continue working on a more general
method to monitor errors in the evolution of black hole spacetimes.
Finally, in an independent project, we are developing a version of 
\dendrogr\ that runs primarily on GPUs.

The version of \dendrogr\ used to produce the results in this paper is 
distributed subject to the MIT license at
\href{https://github.com/paralab/Dendro-GR}{https://github.com/paralab/Dendro-GR}.

\begin{acknowledgments}
This research was supported by NASA 80NSSC20K0528 and NSF PHY-1912883 (BYU). Y.Z. was also supported by NSF awards No.\ OAC-2004044, No.\ PHY-2110338, and PHY-2207920.
This work used the Extreme                                                                               
Science and Engineering
Discovery Environment (XSEDE), which is
supported by NSF Grant No. ACI-1548562, and the Frontera Pathways allocation PHY22019.
Additional computational resources were provided by the GreenPrairies and WhiteLagoon clusters at the Rochester Institute of Technology, which were  supported by NSF grants No.\ PHY-2018420 and No.\ PHY-1726215.
\end{acknowledgments}

\clearpage

\bibliography{paper}{} 

\end{document}